# An Ensemble Kushner-Stratonovich (EnKS) Nonlinear Filter: Additive Particle Updates in Non-Iterative and Iterative Forms

Saikat Sarkar and Debasish Roy*

Computational Mechanics Lab, Department of Civil Engineering, Indian Institute of Science, Bangalore 560012, India

*Corresponding author; email: royd@civil.iisc.ernet.in

**Abstract:**

Despite the cheap availability of computing resources enabling faster Monte Carlo simulations, the potential benefits of particle filtering in revealing accurate statistical information on the imprecisely known model parameters or modeling errors of dynamical systems, based on limited time series data, have not been quite realized. A major numerical bottleneck precipitating this under-performance, especially for higher dimensional systems, is the progressive particle impoverishment owing to weight collapse and the aim of the current work is to address this problem by replacing weight-based updates through additive ones. Thus, in the context of nonlinear filtering problems, a novel additive particle update scheme, in its non-iterative and iterative forms, is proposed based on manipulations of the innovation integral in the governing Kushner-Stratonovich equation. Numerical evidence for the identification of nonlinear and large dimensional dynamical systems indicates a substantively superior performance of the non-iterative version of the EnKS vis-à-vis most existing filters. The costlier iterative version, though conceptually elegant, mostly appears to effect a marginal improvement in the reconstruction accuracy over its non-iterative counterpart. Prominent in the reported numerical comparisons are variants of the Ensemble Kalman Filter (EnKF) that also use additive updates, albeit with many inherent limitations of a Kalman filter.

**Keywords:** *Kushner-Stratonovich equation; Euler approximation; inner iterations; Monte Carlo filters; error estimates; nonlinear system identification*

# 1. Introduction

In recent years, stochastic filters based on Monte Carlo (MC) simulations have gained prominence owing to their potential in solving a large class of nonlinear estimation problems, ranging from dynamical state/parameter estimation to complex target tracking, atmospheric data assimilation etc., by combining partially observed noisy time series data acquired through experiments with the simulation tools for the dynamical system model. The original filter by Kalman and Bucy [1] may be traced back as a progenitor of most modern stochastic filters, even though the former was developed as a closed-form analytical (and not as an MC) scheme for solving strictly linear estimation problems with Gaussian noises. In stochastic filtering, the estimation problem is posed as determining the distributions of (measurable functions of) the system states, also called processes, conditioned on the filtration generated by the measurements up to the current time $t$. When the drift terms in the stochastic differential equations (SDE-s) representing the process and measurement dynamics are linear and the noises or diffusion terms Gaussian, the evolving conditional distribution for the estimation problem, also called the filtering distribution, is Gaussian which is determined by the evolution equations for mean and covariance, the first two moments, as prescribed by the Kalman filter [2]. Even though the idea behind this filter is simple and elegant, its sub-optimal extensions based on drift linearization, the extended Kalman filters (EKF-s) to wit [3], are often found inadequate in treating estimation problems with strong drift nonlinearity and/or noise non-Gaussianity in the process/measurement dynamics. The last class of problems typically involve non-Gaussian, possibly multi-modal conditional distributions as solutions to the estimation problems. Here a MC simulation approach, adopted by many recent filters such as the particle filters (PF-s), is preferred as evolutions of such distributions cannot generally be described by a finite hierarchical set of moments [4] except for a few special cases. For nonlinear, non-Gaussian estimation problems, the governing filtering equations describing the evolution of the conditional distribution in the space of probability measures may be derived starting with an appropriate change of measures, which immediately yields the Kallianpur-Striebel formula [5]. Manipulations of this formula based on the principles of stochastic calculus leads to the Zakai equation [6] for the un-normalized conditional density and onwards to the Kushner-Stratonovich (KS) equation [7] for the normalized density. Unfortunately, analytical or even acceptably accurate numerical solutions to these equations are not available [8]. During the early stages of development of MC filters approximating the nonlinear filtering equations during the last quarter of the last century, success was limited partly owing to a rather restricted availability of the computational power. This probably led to a somewhat popular adoption of the analytical EKF for nonlinear filtering applications [9], even though the method was known to be potentially inaccurate and often unstable thereby requiring an elaborate tuning of the process noise covariance.

The vastly improved computational power over the last couple of decades has imparted a great fillip to the devising of innovative numerical schemes that solve the non-linear and non-Gaussian filtering problem. Amongst these, prominence may be accorded to sequential Monte Carlo (SMC) techniques, i.e. the PF-s [10-14], which are basically recursive Bayesian approaches empirically representing the posterior (or filtering) distribution through an ensemble of MC realizations of the system states, also called particles. Extensive convergence studies have been carried out for many of these numerical schemes which show that these schemes, through multiplicative weight-based updates, empirically represent the desired filtering distribution at a given time with the error decreasing (in law) proportional to only $\frac{1}{\sqrt{N}}$, where $N$ denotes the number of particles or the ensemble size [15, 16]. Unfortunately, most SMC techniques are scourged with the problem of 'particle impoverishment', especially when applied to higher dimensional filtering problems wherein the weights tend to collapse sequentially to a point mass. Once this happens, the process upon conditioning on the measurement till the current time receives no non-trivial updates. Numerical evidence suggests that the typical ensemble size preventing 'weight collapse' increases exponentially with increasing system dimension [16]. Among the numerous research articles aiming at improving these SMC techniques, implicit sampling [17], improved re-sampling [18] and Markov chain Monte Carlo (MCMC) sampling based particle filters [19] have, amongst others, drawn attention. Since methods like implicit sampling, improved re-sampling etc. still rely on the multiplicative weight based update strategy, they are not essentially free from the problem of 'particle impoverishment'. On the other hand, MCMC schemes typically take a very large number of iterations for most problems of practical interest and may thus be computationally unwieldy. Within the MC setup, the numerical infeasibility of requiring impractically large $N$ is, to an extent, bypassed by an MC based filter called the ensemble Kalman filter (EnKF) [20] and its variants, several of which have been successfully applied to large dimensional atmospheric data assimilation problems. Here a resolution to the problem of particle impoverishment is realized through an additive update, which is basically an MC implementation of the gain-based update term of the Kalman filter. But a major criticism of this class of approximate schemes is that they are derived heuristically and even though drift non-linearity is accounted for in some way, they are hardly equipped to treat noise non-Gaussianity [20]. There are attempts in the literature to directly solve the Zakai equation by approximating the un-normalized filtering distribution via time and space discretizations or through functional series [21-23]. For example, the conditional density is approximated using multiple Wiener and Stratonovich integrals in [21,24]. In [22,23], numerical approximations to the Zakai equation have been validated through low-dimensional problems. Even though the Zakai equation is linear and widely studied, in numerical computations it suffers from serious deficiencies [8], e.g. fast dissipation of the solution with increased time step and intermittency leading to rare but large peaks. A way of circumventing these numerical limitations would be to take recourse to the KS equation [8].

The KS equation, the parent filtering equation derivable through Ito's expansion of the Kallianpur-Striebel formula, gives the evolution of the normalized conditional distribution (or a measure-valued conditional process) via a stochastic integral expression. But, except for a few very special cases of linearity and Gaussianity, the KS equation cannot, in general, be reduced to a closed set of stochastic partial differential equations (SPDEs), which can be numerically integrated to arrive at the desired filtering distribution at a given time. Indeed a direct approximation of the KS equation, say using an Euler-type discretization, does not generally yield an accurate and robust scheme. Many SMC methods, which attempt MC simulations based on averaging over the characteristics (i.e. the sample paths provided by the process and measurement dynamics) following a conditional Feynman-Kac formula [25] and typically leading to a weighted particle system [26], have been tried to approximate the KS equation. However, as noted before, most of these methods are not free from the curse of weight collapse, even for moderately large filter dimensions. The primary aim of this article is a resolution of this long standing problem through an efficient, yet accurate, additive particle update scheme derived through the KS equation.

Building upon our recent idea of an MC-based iterative approximation to the additive update term in the KS equation [27], we propose a substantively modified and improved version of the filter that, whilst closely following the evolutions of the estimates based on the KS equation, efficiently implements the nonlinear and strictly additive particle updates without an imperative necessity for inner iterations. This development is based on a sequence of manipulations of the update term so as to introduce an additional layer of numerical dispersion in the gain-like coefficient of the innovation. While inner iterations, as in the previous version of the filter, may still be utilized with some improvement in the estimate, the non-iterative form of the filter does yield solutions that are quite accurate even for large dimensional nonlinear filtering problems. Proofs of convergence of the modified filter, in both its non-iterative and iterative forms, are also provided.

The rest of the paper is organized as follows. Section 2 introduces the filtering problem in a generic form. In Section 3, discretization of the KS equation is discussed. The two versions of the proposed filter, non-iterative and iterative, are detailed (along with pseudo-codes) in Sections 4 and 5 respectively. Numerical illustrations are provided in Section 6 and, finally, the concluding remarks given in Section 7. Proofs of **Theorems 1, 2 and 3** are appear in **Appendix I**.

## 2. Statement of the problem

Within a complete probability space $(\Omega, \mathcal{F}, P)$, supplied with an increasing filtration $\{\mathcal{F}_t, 0 \leq t \leq T_{\max}\}$ consisting of $\sigma$-subalgebras of $\mathcal{F}$, the system process model, typically represented as Ito stochastic differential equations (SDE-s), has the generic form

$$dX_t = b(X_t, t)dt + f(X_t, t)dB_t \tag{2.1}$$

for $t \in (t_{i-1}, t_i]$, $i=1,2,3,\ldots$ with $X_t := X(t) \in \mathbb{R}^n$ the (hidden) process vector, $b: \mathbb{R}^n \times \mathbb{R}_+ \mapsto \mathbb{R}^n$ the non-linear drift function, $f: \mathbb{R}^n \times \mathbb{R}_+ \mapsto \mathbb{R}^{n \times m}$ the diffusion matrix and $B_t \in \mathbb{R}^m$ an $m$-dimensional standard $P$-Brownian motion. Here $0 = t_0 < t_1 \ldots < t_i < \ldots < t_N = T_{\max}$ denotes an ordering of the time interval $(0, T_{\max}]$ of interest to facilitate a recursive numerical implementation of the stochastic filtering scheme. In contrast to the system process model, the measurement model is available typically in the algebraic form

$$Y_t = h(X_t,t) + \Delta \eta_t \qquad (2.2a)$$

where $Y_t := Y(t) \in \mathbb{R}^q$ ($y_0 := 0$) is the noisy measurement process generating the sub-filtration $\mathcal{F}_t^Y$. $h: \mathbb{R}^n \times \mathbb{R}_+ \mapsto \mathbb{R}^q$ is the non-linear measurement function and $\Delta \eta_t = \eta_t - \eta_{i-1} \in \mathbb{R}^q$ a $q$-dimensional $P$-Brownian increment representing the measurement noise (assuming that the last measurement $Y_{i-1}$ was sampled at $t_{i-1}$). An incremental form of the noise term $\Delta \eta_t$ is used in Eqn. (2.2a) so as to indicate its relative 'smallness' vis-à-vis $h(X_t,t)$, the measurement drift function. Since the nonlinear filtering (KS) equation demands SDE structures for both the system process and measurement models, we define a fictitious process $Y_t := Y(t)$ (with $Y_0 := 0$) so that the measurement equation over $(t_{i-1}, t_i]$ in terms of incremental $Y_t$ is given by

$$\Delta Y_t := Y_t \Delta t = h(X_t,t)\Delta t + \Delta \eta_t \Delta t; \quad \Delta t = t - t_{i-1} \qquad (2.2b)$$

Assuming the time step $\Delta t_i := t_i - t_{i-1}$ to be small, the above equation may be approximated as

$\Delta Y_t \approx h(X_t,t)\Delta t + \Delta \eta_t \Delta t_i$, which corresponds to the SDE

$$dY_t \approx h(X_t,t)dt + \Delta t_i d\eta_t \qquad (2.2c)$$

Note that the term $\Delta t \Delta \eta_t$ is of $\mathcal{O}(\Delta t^{3/2})$ vis-à-vis the drift term $h(X_t,t)\Delta t \sim \mathcal{O}(\Delta t)$. The above representation of the noise term is in keeping with the general scenario of the measurement noise being smaller than the measurement drift term by an order of magnitude in the mean square sense. The relative 'smallness' of the noise term is justifiable through the fact that most modern sensing devices used for data acquisition are very accurate with high signal-to-noise ratio. We now write $d\eta_t$ in terms of a standard $P$-Brownian increment $dW_t \in \mathbb{R}^q$ as $d\eta_t = \upsilon_t dW_t$ where $\upsilon_t := \upsilon(t)$ is a $q \times q$ dimensional matrix representing the measurement noise intensity. Defining $\boldsymbol{\sigma}_t = \upsilon_t \Delta t_i$ (the scaled measurement noise intensity), the measurement SDE is of the final form:

$$dY_t = h(X_t,t)dt + \boldsymbol{\sigma}_t dW_t \qquad (2.3)$$

It is emphasized that, in implementing the new EnKS filter, $|\sigma_t|$ need not actually be 'small' and that the noise term need not be Gaussian. Thus the measurement diffusion matrix could be of the form $\sigma_t = \sigma(X_t, t)$, a function of $X_t$. While the presentation to follow could also be adapted to non-Brownian (e.g. Poisson's) right continuous noise processes (possibly with identically zero quadratic co-variation), we presently stick to the measurement SDE (2.3) so that we have a strictly continuous form of measurement filtration $\mathcal{F}_{t+}^Y = \mathcal{F}_t^Y$. Standard existence criteria [28] for weak solutions to the above SDEs are assumed. The purpose of stochastic filtering is then to arrive at the conditional (filtered) distribution of, say, a scalar-valued function $\phi(X_t)$, $\phi \in C_b^2$ (bounded and twice continuously differentiable), given the measurement history $\mathcal{F}_t^Y := \sigma\{Y_s \mid s \in (0,t]\}$. Thus the conditional estimate $\pi_t(\phi)$ is defined as the measure-valued process $E_P(\phi(X_t) \mid \mathcal{F}_t^Y)$ measurable with respect to $\mathcal{F}_t^Y$.

## 3. Discretization of the KS equation

The filtered estimate $\pi_t(\phi)$ of $\phi(X_t)$, for $t \in (t_{i-1}, t_i]$, satisfies the KS equation (or the filtered martingale problem [29]):

$$\pi_t(\phi) = \pi_{i-1}(\phi) + \int_{t_{i-1}}^{t} \pi_s(L_s(\phi)) ds + \sum_{\varsigma=1}^{q} \int_{t_{i-1}}^{t} \left\{ \pi_s(M_s^\varsigma(\phi)) - \pi_s(h^\varsigma(\cdot, s)) \pi_s(\phi) \right\} dI_s^\varsigma \qquad (3.1)$$

$dI_t := \{dI_t^\varsigma\} = (\sigma_t^T \sigma_t)^{-1} \{dY_t - \pi_t(h) dt\}$ denotes the incremental innovation vector process and $I_t^\varsigma$ the $\varsigma^{th}$ element of $I_t$. Here $\pi_{i-1}(\cdot) := \pi_{t_{i-1}}(\cdot)$ and $L_t$ is the infinitesimal generator corresponding to the system process SDE (2.1) given by

$$L_t(\phi(x)) := L(\phi(x_t))$$
$$= \frac{1}{2} \sum_{\xi=1}^{n} \sum_{\eta=1}^{n} a^{\xi\eta}(x_t, t) \frac{\partial^2 \phi(x_t)}{\partial x_t^\xi \partial x_t^\eta} + \sum_{\xi=1}^{n} b^\xi(x_t, t) \frac{\partial \phi(x_t)}{\partial x_t^\xi}, \quad x_t = \{x_t^1, \ldots, x_t^n\}^T \in \mathbb{R}^n$$

$a := ff^T$ with $a^{\xi\eta}$ denoting the $(\xi,\eta)^{\text{th}}$ element of the matrix $a$. Similarly, $b^{\xi}$ is the $\xi^{\text{th}}$ element of the vector $b$. Moreover, we have $\mathrm{M}_t^{\varsigma}(\phi(x)) := \mathrm{M}^{\varsigma}(\phi(x_t)) = h^{\varsigma}(x_t,t)\phi(x_t)$, where $h^{\varsigma}$, $Y_t^{\varsigma}$ are the $\varsigma^{\text{th}}$ elements of the vectors $h$ and $Y_t$ respectively. Indeed, Eqn. (3.1) may be interpreted as a weak form to determine the conditional measure $\pi_t(\cdot)$ with $\phi$ being a test function. The aim of a typical filtering method is to design a numerical scheme so that the innovation process is driven to a zero-mean martingale (corresponding to the diffusion term of the measurement SDE 2.3) through recursion over time $t$. This goal is often accomplished in two major stages, viz. prediction and update. In most MC filters, the prediction stage involves integrating the system process SDE (2.1) over $(t_{i-1}, t_i]$ starting with the realizations (particles) of the filtered solution at $t_{i-1}$ as the initial conditions and hence this stage is often executed independent of (i.e. as a precursor to) the update stage. A similar strategy is adopted in developing the current filtering scheme as well. A first step in this direction would be to approximate the second term on the right hand side (RHS) of the KS Eqn. (3.1) as

$$\int_{t_{i-1}}^{t} \pi_s(\mathrm{L}_s(\phi))\,ds \cong \pi_{i-1} \int_{t_{i-1}}^{t} \mathrm{L}_s(\phi)\,ds \tag{3.1a}$$

Recall that Eqn. (3.1) is arrived at after averaging over the diffusion paths corresponding to the process noise $B_t$. Moreover, if the second term on the RHS of the KS Eqn. (3.1) is replaced by the approximation in Eqn. (3.1a), then the first two terms (referred to as 'the prediction component') on the RHS of the KS equation, so approximated, recover Dynkin's formula for the predicted mean $E_P(\phi(X_t) | X(t_{i-1}) := X_{i-1})$ according to the process dynamics of Eqn. (1.1). By way of motivating the EnKS filter, a particle based representation (the unmasked form) of Eqn. (3.1) may be conceived of by putting back, in the prediction component, the diffusion term for the process dynamics. As a first step in deriving the unmasked additive update, as the current measurement $Y_t$ (typically at $t = t_i$) is available, an MC setting for solving Eqn. (3.1) may be set up as

$$\pi'_t(\phi) = \pi'_{i-1}(\phi) + \int_{t_{i-1}}^{t} \pi'_s(\mathrm{L}(\phi))\,dt + \sum_{\varsigma=1}^{q} \int_{t_{i-1}}^{t} \left\{ \pi'_s(h^{\varsigma}\phi) - \pi'_s(h^{\varsigma})\pi'_s(\phi) \right\} dI'^{\varsigma}_s \tag{3.2}$$

where $dI'_t := \{dI'^\varsigma_t\} = (\boldsymbol{\sigma}_t^T \boldsymbol{\sigma}_t)^{-1} \{dY_t - \pi'_t(h)dt\}$, $\pi'(\phi) = (1/N)\sum_{j=1}^{N} \phi^{(j)}$ is the ensemble-averaged approximation of the actual conditional estimate. Note that the bracketed superscript $^{(j)}$ over a random variable represents its $j^{\text{th}}$ realization and that $\phi^{(j)} := \phi(X^{(j)})$. Let $\boldsymbol{\varphi} := (\phi^1,...,\phi^{\hat{n}})^T$ be an $\hat{n}$-dimensional vector-valued function to be estimated via the filtering technique with $\phi^k(x_t) \in C_b^2$ for $k \in [1,\hat{n}]$; typically we have $\boldsymbol{\varphi}(x) := (x^1,...,x^n)^T$ so that $\hat{n} = n$ and $\phi^k(x) = x^k$. Then an unmasked (particle based) representation of Eqn. (3.2) may be written as

$$\boldsymbol{\Phi}_t = \boldsymbol{\Phi}_{i-1} + \int_{t_{i-1}}^{t} \boldsymbol{\Psi}_s ds + \int_{t_{i-1}}^{t} d\boldsymbol{\Xi}_s + \frac{1}{N}\int_{t_{i-1}}^{t} \left\{\boldsymbol{\Phi}_s \mathbf{H}_s^T - \widehat{\boldsymbol{\Phi}}_s \widehat{\mathbf{H}}_s^T\right\}\left(\boldsymbol{\sigma}_s^T \boldsymbol{\sigma}_s\right)^{-1}\{d\mathbf{Y}_s - \mathbf{H}_s ds\} \quad (3.3)$$

where $\boldsymbol{\Phi}_t := [\boldsymbol{\varphi}_t^{(1)},...\boldsymbol{\varphi}_t^{(N)}]$, $\boldsymbol{\Phi}_{i-1} := \boldsymbol{\Phi}_{t_{i-1}}$, $\boldsymbol{\Psi}_t := [\mathrm{L}(\boldsymbol{\varphi}_t^{(1)}),...,\mathrm{L}(\boldsymbol{\varphi}_t^{(N)})]$,

$d\boldsymbol{\Xi}_t := [\boldsymbol{\varphi}_t^{\prime(1)} f_t^{(1)} dB_t^{(1)},...,\boldsymbol{\varphi}_t^{\prime(N)} f_t^{(N)} dB_t^{(N)}]$, $\mathbf{H}_t := [h_t^{(1)},...,h_t^{(N)}]$, $\widehat{\boldsymbol{\Phi}}_t = [\pi'_t(\boldsymbol{\varphi}),...,\pi'_t(\boldsymbol{\varphi})] \in \mathbb{R}^{n \times N}$,

$\widehat{\mathbf{H}}_t = [\pi'_t(h),...,\pi'_t(h)] \in \mathbb{R}^{q \times N}$

and $d\mathbf{Y}_t := [dY_t,...,dY_t] \in \mathbb{R}^{q \times N}$. Note that the identical (column) vector elements of $\widehat{\boldsymbol{\Phi}}_t$ and $\widehat{\mathbf{H}}_t$ are respectively the ensemble-averaged $\boldsymbol{\varphi}$ and $h$ respectively at the current time $t$. As a precursor to the rest of the derivation, it is convenient to recast Eqn. (3.3) as:

$$\boldsymbol{\Phi}_t = \boldsymbol{\Phi}_{i-1} + \int_{t_{i-1}}^{t} (\boldsymbol{\Psi}_s ds + d\boldsymbol{\Xi}_s) + \frac{1}{N}\int_{t_{i-1}}^{t} \left\{\boldsymbol{\Phi}_s \mathbf{H}_s^T - \widehat{\boldsymbol{\Phi}}_s \mathbf{H}_s^T + \widehat{\boldsymbol{\Phi}}_s \mathbf{H}_s^T - \widehat{\boldsymbol{\Phi}}_s \widehat{\mathbf{H}}_s^T\right\}\left(\boldsymbol{\sigma}_s^T \boldsymbol{\sigma}_s\right)^{-1}\{d\mathbf{Y}_s - \mathbf{H}_s ds\}$$

(3.4)

A major hindrance in using Eqn. (3.4), an MC based unmasked representation of Eqn. (3.1), continues to be the problem of circularity in that the expectation of $\boldsymbol{\Phi}_t$ needs information on that of $\boldsymbol{\Phi}_t \mathbf{H}_t^T$, i.e. higher order expectations. This impedes a direct solution of the set of non-linear equations. Starting with Eqn. (3.4), an algorithm is devised in the next section to circumvent this problem.

## 4. The EnKS methodology: a non-iterative form

Within the MC-setting used to address the problem of moment closure in Eqn. (3.4), a two-stage strategy is adopted in the proposed EnKS filter. First, for a given $t \in (t_{i-1}, t_i]$ in the current time interval, the process SDE (corresponding to the first three terms on the RHS of Eqn. (3.4)) is weakly solved using a numerical integration technique (e.g. Euler Maruyama (EM) [30], Milstein's scheme, local linearization [31, 32, 33] or stochastic Newmark [34] schemes etc.). In the second stage, an MC-based additive update term is derived using the fourth term on the RHS of Eqn. (3.4). Although an explicit EM scheme is considered here for numerical integration of the process SDEs, a more accurate/stable stochastic integration scheme could be adopted to enhance the reconstruction fidelity. Presently, using the explicit EM-based integration, the recursive prediction-update filtering strategy that aims at arriving at an empirical filtered distribution at time $t \in (t_{i-1}, t_i]$ is depicted below. In all the numerical work, however, we set $t = t_i$.

*Prediction*

$$\tilde{\boldsymbol{\Phi}}_t = \boldsymbol{\Phi}_{i-1} + \boldsymbol{\Psi}_{i-1} \Delta t + \Delta \boldsymbol{\Xi}_{i-1} \tag{4.1}$$

where, $\boldsymbol{\Psi}_{i-1} := \boldsymbol{\Psi}_{t_{i-1}}$, $\Delta t = t - t_{i-1}$

$\Delta \boldsymbol{\Xi}_{i-1} := \Delta \boldsymbol{\Xi}_{t_{i-1}} = [\varphi_t'^{(1)} f_{i-1}^{(1)} \left( B_i^{(1)} - B_{i-1}^{(1)} \right), ..., \varphi_t'^{(N)} f_{i-1}^{(N)} \left( B_i^{(N)} - B_{i-1}^{(N)} \right)]$, $f_{i-1} := f_{t_{i-1}}$, and $B_{i-1} := B_{t_{i-1}}$.

*Additive Update*

A time-discretized MC-form of the update equation, motivated by an EM-based approximation to the third term on the RHS of Eqn. (3.4), may be written as:

$$\boldsymbol{\Phi}_t = \tilde{\boldsymbol{\Phi}}_t + \frac{1}{N} \left\{ \left( \tilde{\boldsymbol{\Phi}}_t - \hat{\tilde{\boldsymbol{\Phi}}}_t \right) \tilde{\mathbf{H}}_t^T + \hat{\tilde{\boldsymbol{\Phi}}}_t \left( \tilde{\mathbf{H}}_t^T - \hat{\tilde{\mathbf{H}}}_t^T \right) \right\} \left( \boldsymbol{\sigma}_t^T \boldsymbol{\sigma}_t \right)^{-1} \left\{ \Delta \mathbf{Y}_t - \tilde{\mathbf{H}}_t \Delta t \right\} \tag{4.2a}$$

Here $\tilde{\mathbf{H}}_t := [\tilde{h}_t^{(1)}, ..., \tilde{h}_t^{(N)}] = [h^{(1)}(\tilde{X}_t), ..., h^{(N)}(\tilde{X}_t)]$ is the predicted ensemble of the measurement drift vectors. Recalling from Eqn. (2.2b) that $\Delta Y_t := Y_t \Delta t$, Eqn. (4.2a) may be recast as:

$$\boldsymbol{\Phi}_t = \tilde{\boldsymbol{\Phi}}_t + \frac{1}{N}\left\{\left(\tilde{\boldsymbol{\Phi}}_t - \hat{\tilde{\boldsymbol{\Phi}}}_t\right)\left(\tilde{\mathbf{H}}_t^T \Delta t\right) + \left(\hat{\tilde{\boldsymbol{\Phi}}}_t \Delta t\right)\left(\tilde{\mathbf{H}}_t^T - \hat{\tilde{\mathbf{H}}}_t^T\right)\right\}\left(\boldsymbol{\sigma}_t^T \boldsymbol{\sigma}_t\right)^{-1}\left\{\hat{\mathbf{Y}}_t - \tilde{\mathbf{H}}_t\right\} \qquad (4.2b)$$

where $\hat{\mathbf{Y}}_t = [Y_t, ..., Y_t] \in \mathbb{R}^{q \times N}$. Recall that while $\tilde{\boldsymbol{\Phi}}_t = [\tilde{\boldsymbol{\varphi}}_t^{(1)}, ..., \tilde{\boldsymbol{\varphi}}_t^{(N)}]$ contains the predicted particles within the ensemble, $\hat{\tilde{\boldsymbol{\Phi}}}_t$ is constructed using $N$ identical column vectors of the ensemble-averaged prediction $\pi_t'(\tilde{\boldsymbol{\varphi}})$. In the initial stages of time evolution, when the innovation process could have a significant drift component owing to the measurement-prediction mismatch (i.e. a significant departure from a zero-mean martingale), the gain-type coefficient matrix should be such (e.g. having a large norm) that the sample space is better explored. Keeping in mind that the temporal gradients of the evolving estimates have sharper gradients in this regime, one way to construct the gain-type coefficient matrix would be to incorporate information on these gradients through the previous estimates. Thus $\tilde{\mathbf{H}}_t \Delta t$ and $\hat{\tilde{\boldsymbol{\Phi}}}_t \Delta t$ may be approximated as:

$$\tilde{\mathbf{H}}_t \Delta t \approx \left(\tilde{\mathbf{H}}_t t - \hat{\tilde{\mathbf{H}}}_{i-1} t_{i-1} - \Delta \hat{\tilde{\mathbf{H}}}_t t\right) \qquad (4.3a)$$

$$\hat{\tilde{\boldsymbol{\Phi}}}_t \Delta t = \left(\hat{\tilde{\boldsymbol{\Phi}}}_t t - \hat{\tilde{\boldsymbol{\Phi}}}_{i-1} t_{i-1}\right) \qquad (4.3b)$$

where Ito's formula has been used in writing the approximation (4.3a). Since this modification incorporates the filtered estimates at the previous time instant ($t_{i-1}$), it should be helpful in expediting filter convergence especially in regions of sample space far away from the converged solution. Thus we get;

$$\boldsymbol{\Phi}_t = \tilde{\boldsymbol{\Phi}}_t + \frac{1}{N}\left\{\begin{array}{l}\left(\tilde{\boldsymbol{\Phi}}_t - \hat{\tilde{\boldsymbol{\Phi}}}_t\right)\left(\tilde{\mathbf{H}}_t^T t - \hat{\tilde{\mathbf{H}}}_{i-1}^T t_{i-1} - \Delta \hat{\tilde{\mathbf{H}}}_t^T t\right) \\ + \left(\hat{\tilde{\boldsymbol{\Phi}}}_t t - \hat{\tilde{\boldsymbol{\Phi}}}_{i-1} t_{i-1} - \Delta \hat{\tilde{\boldsymbol{\Phi}}}_t t\right)\left(\tilde{\mathbf{H}}_t^T - \hat{\tilde{\mathbf{H}}}_t^T\right)\end{array}\right\}\left(\boldsymbol{\sigma}_t^T \boldsymbol{\sigma}_t\right)^{-1}\left\{\hat{\mathbf{Y}}_t - \tilde{\mathbf{H}}_t\right\} \qquad (4.4)$$

It may now be observed that, once the converged filtered estimate is available, the squared noise intensity term $\boldsymbol{\sigma}_t^T \boldsymbol{\sigma}_t$ may be replaced by the innovation covariance matrix $E_P\left(\left((Y_t - h_t) - \pi_t'(Y_t - h_t)\right)\left((Y_t - h_t) - \pi_t'(Y_t - h_t)\right)^T\right)$. However, away from the converged solution and especially during the initial stages of the filter evolution, the last covariance matrix

(e.g. its norm) would typically be rather 'large'. Incorporation of this term, in lieu of $\boldsymbol{\sigma}_t^T \boldsymbol{\sigma}_t$ in Eqn. (4.4) would have the effect of artificially increasing the measurement noise in the initial stages, thereby enabling the diffusion term in the system process model to efficaciously explore the search space. In the MC setup that we have adopted, the innovation covariance matrix may be computed as:

$$E_P\left(\left((Y_t - h_t) - \pi_t'(Y_t - h_t)\right)\left((Y_t - h_t) - \pi_t'(Y_t - h_t)\right)^T\right) = \frac{1}{N-1}\left(\tilde{\mathbf{H}}_t^T - \hat{\tilde{\mathbf{H}}}_t^T\right)\left(\tilde{\mathbf{H}}_t - \hat{\tilde{\mathbf{H}}}_t\right)$$

Finally, introducing a scalar parameter $0 < \alpha < 1$, $\left(\boldsymbol{\sigma}_t^T \boldsymbol{\sigma}_t\right)^{-1}$ in Eqn. (4.4) is replaced by

$$\left(\alpha \frac{1}{N-1}\left(\tilde{\mathbf{H}}_t^T - \hat{\tilde{\mathbf{H}}}_t^T\right)\left(\tilde{\mathbf{H}}_t - \hat{\tilde{\mathbf{H}}}_t\right) + (1-\alpha)\left(\boldsymbol{\sigma}_t^T \boldsymbol{\sigma}_t\right)\right)^{-1}$$

Eqn. (4.4) thus takes the form:

$$\boldsymbol{\Phi}_t = \tilde{\boldsymbol{\Phi}}_t + \frac{1}{N}\left\{\left(\tilde{\boldsymbol{\Phi}}_t - \hat{\tilde{\boldsymbol{\Phi}}}_t\right)\left(\tilde{\mathbf{H}}_t^T t - \hat{\tilde{\mathbf{H}}}_{i-1}^T t_{i-1} - \Delta \hat{\tilde{\mathbf{H}}}_t^T t\right) + \left(\hat{\tilde{\boldsymbol{\Phi}}}_t t - \hat{\tilde{\boldsymbol{\Phi}}}_{i-1} t_{i-1}\right)\left(\tilde{\mathbf{H}}_t^T - \hat{\tilde{\mathbf{H}}}_t^T\right)\right\}$$

$$\left\{\alpha \frac{1}{N-1}\left(\tilde{\mathbf{H}}_t - \hat{\tilde{\mathbf{H}}}_t\right)\left(\tilde{\mathbf{H}}_t^T - \hat{\tilde{\mathbf{H}}}_t^T\right) + (1-\alpha)\boldsymbol{\sigma}_t^T \boldsymbol{\sigma}_t\right\}^{-1}\left\{\hat{\mathbf{Y}}_t - \tilde{\mathbf{H}}_t\right\}$$

(4.5)

A more concise form of the update equation may be written as:

$$\boldsymbol{\Phi}_t = \tilde{\boldsymbol{\Phi}}_t + \tilde{\mathbf{G}}_t\left\{\hat{\mathbf{Y}}_t - \tilde{\mathbf{H}}_t\right\} \tag{4.6}$$

where

$$\tilde{\mathbf{G}}_t := \frac{1}{N}\left\{\left(\tilde{\boldsymbol{\Phi}}_t - \hat{\tilde{\boldsymbol{\Phi}}}_t\right)\left(\tilde{\mathbf{H}}_t^T t - \hat{\tilde{\mathbf{H}}}_{i-1}^T t_{i-1} - \Delta \hat{\tilde{\mathbf{H}}}_t^T t\right) + \left(\hat{\tilde{\boldsymbol{\Phi}}}_t t - \hat{\tilde{\boldsymbol{\Phi}}}_{i-1} t_{i-1}\right)\left(\tilde{\mathbf{H}}_t^T - \hat{\tilde{\mathbf{H}}}_t^T\right)\right\}$$

$$\left\{\alpha \frac{1}{N-1}\left(\tilde{\mathbf{H}}_t - \hat{\tilde{\mathbf{H}}}_t\right)\left(\tilde{\mathbf{H}}_t^T - \hat{\tilde{\mathbf{H}}}_t^T\right) + (1-\alpha)\boldsymbol{\sigma}_t^T \boldsymbol{\sigma}_t\right\}^{-1} \quad .$$

In line with the traditional stochastic filtering terminology (e.g. the Kalman filter), the update term may be thought of as an innovation term $\mathbf{I}_t := \{\widehat{\mathbf{Y}}_t - \widetilde{\mathbf{H}}_t\}$, weighted by the gain-type coefficient matrix $\widetilde{\mathbf{G}}_t$. A pseudo-code for the non-iterative EnKS is provided below.

*Pseudo-code 1: for the non-iterative EnKS*

1. Discretize the time interval of interest, say $[0,T]$, using a partition $\{t_0, t_1, ..., t_M\}$ such that $0 = t_0 < t_1 < ... < t_M = T$ and $t_i - t_{i-1} = \Delta t_i$ ($= \dfrac{1}{M}$ if the step size is chosen uniformly for $i = 0, ..., M-1$). Choose an ensemble size $N$.

2. Generate the ensemble of initial conditions $\{\boldsymbol{\varphi}_0^{(j)}\}_{j=1}^N$, or equivalently $\{X_0^{(j)}\}_{j=1}^N$, for the system state vector. For each discrete time instant $t_i, i = 1, ..., M-1$, execute the following steps.

3. *Prediction*

   Using $\{\boldsymbol{\varphi}_{i-1}^{(j)}\}_{j=1}^N$, the update available at the last time instant $t_{i-1}$, propagate each particle to the current time instant $t_i$ using any appropriate integration scheme for SDEs, e.g. an explicit Euler-Maruyama (EM) approximation to Eqn. (2.1) given by:

   $$\tilde{\boldsymbol{\varphi}}_i^{(j)} = \boldsymbol{\varphi}_{i-1}^{(j)} + \mathrm{L}(\boldsymbol{\varphi}_{i-1}^{(j)})\Delta t_i + \boldsymbol{\varphi}_{i-1}^{\prime(j)} f_{i-1}^{(j)} \left( B_i^{(N)} - B_{i-1}^{(j)} \right), \quad j = 1, ..., N$$

   Using $\{\tilde{\boldsymbol{\varphi}}_i^{(j)}\}_{j=1}^N$ compute $\{\tilde{X}_i^{(j)}\}_{j=1}^N$. This step is trivial if $\varphi$ is the identity function $\varphi(X) = X$.

   Using $\{\tilde{X}_i^{(j)}\}_{j=1}^N$, compute $\{\tilde{h}_i^{(j)}\}_{j=1}^N = \{h(\tilde{X}_i^{(j)})\}_{j=1}^N$.

   Construct $\tilde{\boldsymbol{\Phi}}_i := [\tilde{\boldsymbol{\varphi}}_i^{(1)}, ... \tilde{\boldsymbol{\varphi}}_i^{(N)}]$, $\widehat{\tilde{\boldsymbol{\Phi}}}_i = [\pi_i'(\tilde{\boldsymbol{\varphi}}), ..., \pi_i'(\tilde{\boldsymbol{\varphi}})]$, $\tilde{\mathbf{H}}_i := [\tilde{h}_i^{(1)}, ..., \tilde{h}_i^{(N)}]$, $\widehat{\tilde{\mathbf{H}}}_i = [\pi_i'(\tilde{h}), ..., \pi_i'(\tilde{h})]$.

4. *Additive update*

   Choose $\alpha \in (0,1)$; a typical prescribed value would be $\alpha \approx 0.8$, even though the method also performs well for other values in the interval indicated.

   Update each particle as

   $$\boldsymbol{\varphi}_i^{(j)} = \tilde{\boldsymbol{\varphi}}_i^{(j)} + \tilde{\mathbf{G}}_i \left\{ Y_i - \tilde{h}_i^{(j)} \right\}, \quad j = 1, \ldots, N, \text{ where}$$

   $$\tilde{\mathbf{G}}_i := \frac{1}{N} \left\{ \left( \tilde{\boldsymbol{\Phi}}_i - \hat{\bar{\boldsymbol{\Phi}}}_i \right) \left( \tilde{\mathbf{H}}_i^T t_i - \hat{\bar{\mathbf{H}}}_{i-1}^T t_{i-1} - \Delta \hat{\bar{\mathbf{H}}}_i^T t_i \right) + \left( \hat{\bar{\boldsymbol{\Phi}}}_i t_i - \hat{\bar{\boldsymbol{\Phi}}}_{i-1} t_{i-1} \right) \left( \tilde{\mathbf{H}}_i^T - \hat{\bar{\mathbf{H}}}_i^T \right) \right\}$$

   $$\left\{ \alpha \frac{1}{N-1} \left( \tilde{\mathbf{H}}_i - \hat{\bar{\mathbf{H}}}_i \right) \left( \tilde{\mathbf{H}}_i^T - \hat{\bar{\mathbf{H}}}_i^T \right) + (1-\alpha) \boldsymbol{\sigma}_i^T \boldsymbol{\sigma}_i \right\}^{-1}$$

5. If $i < M$, go to step 3 with $i = i+1$,

   else terminate the algorithm.

Using a filtered martingale problem setup, the existence and uniqueness of a posterior distribution satisfying the KS equation has been proved in [29] under very general conditions on the drift and diffusion fields of the system process and measurement SDEs. However, since we presently adopt Ito's theory in interpreting the weak solutions of the SDEs, somewhat more restrictive conditions, e.g. Lipschitz continuity and linear growth bound, are applied to the drift and diffusion terms. In order to show that the proposed algorithm converges to the filtered estimate, we have the following theorem (**Theorem 1**).

***Theorem 1***:

Let $\phi \in C_b^2(\mathbb{R})$. Assume that there exist constants $M_1, M_2 > 0$ such that

$$|b(t,x) - b(t,y)| + |f(t,x) - f(t,y)| \leq M_1 |x - y| \tag{4.7}$$

$$|b(t,x)| + |f(t,x)| \leq M_2 |1 + |x|| \tag{4.8}$$

and

$$E_P \left[ |X_0|^2 \right] < \infty \tag{4.9}$$

Assume additionally that $h$ is a bounded and Lipschitz continuous function. Furthermore, we assume that $\phi$ is sufficiently smooth so that $\phi(x)$ and its derivatives satisfy an inequality of the form

$$|\phi(x)| \leq M_3 \left(1 + |x|^a\right) \tag{4.10}$$

for positive constants $M_3, a$. Then there exist constants $D' > 0$ and $D'' > 0$, independent of $\Delta t_i$, such that (we use $\overline{\boldsymbol{\varphi}} := \pi^e(\boldsymbol{\varphi})$, $\overline{h} := \pi^e(h)$ for notational convenience; $\pi^e$ denotes the conditional expectation operator for an EM-discretized argument):

$$\left(E_P\left[\left|\pi(\boldsymbol{\varphi}) - \pi'^e(\boldsymbol{\varphi})\right|^2\right]\right)^{\frac{1}{2}} \leq D'(\Delta t)^{\frac{1}{2}}$$

$$+ \frac{D''}{\sqrt{N}} \Delta t \left(|Y| + \|h\|\right) \left\{ \begin{bmatrix} \left\|L\left((\boldsymbol{\varphi} - \overline{\boldsymbol{\varphi}})(ht - \overline{h}t + \overline{h}_{i-1}\Delta t)^T\right)\right\| \\ + \|\boldsymbol{\varphi}\|\|h\| + \left\|L\left((\overline{\boldsymbol{\varphi}}t - \overline{\boldsymbol{\varphi}}_{i-1}t_{i-1})(h - \overline{h})^T\right)\right\| \end{bmatrix} \left\| \left\{ \begin{matrix} \alpha(h - \overline{h})^T(h - \overline{h}) \\ + (1 - \alpha)\boldsymbol{\sigma}^T\boldsymbol{\sigma} \end{matrix} \right\}^{-1} \right\| \right.$$

$$\left. + \|\boldsymbol{\varphi}\|\|h\|\alpha \left( \Delta t \left\|L\left((h - \overline{h})(h - \overline{h})^T\right)\right\| + \left\|(h - \overline{h})(h - \overline{h})^T\right\| \right) \left\| \left\{ \begin{matrix} \alpha(h - \overline{h})^T(h - \overline{h}) \\ + (1 - \alpha)\boldsymbol{\sigma}^T\boldsymbol{\sigma} \end{matrix} \right\}^{-1} \right\|^2 \right\}$$

$$+ \frac{D''}{\sqrt{N}} \left( \left\|\tilde{\mathbf{G}}(\boldsymbol{\varphi})\right\| \left(\Delta t \|L(h)\| + \|h\|\right) + \left(\Delta t \|L(\boldsymbol{\varphi})\| + \|\boldsymbol{\varphi}\|\right) \right)$$

(4.11)

For both the sake of conciseness and to indicate that the above inequality holds for any $t \in (t_{i-1}, t_i]$, we have removed the subscript '$i$' from $\boldsymbol{\varphi}$, $h$, $t$, $\overline{h}$, $\Delta t$, $\tilde{\mathbf{G}}$ and $\boldsymbol{\sigma}$ in the statement of the theorem. Even in the proof of the theorem given in Appendix I, the subscript '$i$' must be assumed to be present whenever the above variables/operators appear as un-subscripted.

## 5. An iterative version of the EnKS

A motivation in developing an iterative version of the EnKS is derived from the fact that iterations provide an attractive tool for an update procedure involving nonlinearities, e.g. those in the system process and/or measurement models. While the additive nature of particle updates in the EnKS eliminates the curse of 'particle collapse', an iterative form could additionally help precipitate a faster convergence of the measurement-prediction mismatch to a zero-mean martingale. In other words, using an inner layer of iterations, one could attempt a 'maximal' utilization of the current measurement within the particle update before moving over to the next time step. In order to provide an additional boost to the mixing property of the update kernel, an annealing-type scalar sequence of multipliers $\{\beta_0, \beta_1, ..., \beta_{\kappa-1}\}$ is artificially applied to the iterated sequence of gain-weighted innovation terms. Here $\kappa$ denotes the number of inner iterations at the current time $t$. The sequence $\{\beta_0, \beta_1, ..., \beta_{\kappa-1}\}$ is so chosen that $\beta_\kappa \to 1$ as $\kappa \to \infty$ (to approach the original update term) and $\beta_k \leq \beta_{k+1}$ for $k \in [0, \kappa-1)$. Effect of the annealing type term is similar to the temperature (here $1/\beta_k$) in a standard simulated annealing (SA) scheme [35]. The added advantage we have over SA based schemes, is that, unlike SA where a single Markov chain is evolved and $1/\beta_k$ is slowly reduced to unity, in the present setting, an ensemble of pseudo-chains is propagated simultaneously, for a given $t$, allowing us to have a much faster and flexible reduction. Presently an exponential decay, $\beta_{k+1} = \beta_k \exp(k+1-\kappa), k = 0,...,\kappa-1$ which is a more non-conservative schedule, is adopted. At a given time $t$, to initiate the inner iteration, an initial annealing term is required, for which, typically a 'large' value is taken. Although the filtered estimate asymptotically improves with increasing number of $\kappa$, at a given $t$, keeping in mind the computational feasibility, only a few iterations (say, $\kappa = 10$) may be prescribed for a sufficiently large class of problems. With the convention $\{\varphi_{t,0}^{(j)}\}_{j=1}^{N} := \{\tilde{\varphi}_t^{(j)}\}_{j=1}^{N}$ and $\{h_{i,0}^{(j)}\}_{j=1}^{N} := \{\tilde{h}_i^{(j)}\}_{j=1}^{N}$, a generic form of the iterative update may be given as:

$$\boldsymbol{\Phi}_{t,k} = \boldsymbol{\Phi}_{t,k-1} + \beta_{k-1} \mathbf{G}_{t,k-1} \{\hat{\mathbf{Y}}_t - \mathbf{H}_{t,k-1}\}, \quad k = 1,...,\kappa \qquad (5.1)$$

where

$$\mathbf{G}_{t,k} := \frac{1}{N} \{(\boldsymbol{\Phi}_{t,k} - \hat{\boldsymbol{\Phi}}_{t,k})(\mathbf{H}_{t,k}^T t - \hat{\mathbf{H}}_{i-1}^T t_{i-1} - \Delta \hat{\mathbf{H}}_{t,k}^T t) + (\hat{\boldsymbol{\Phi}}_{t,k} t - \hat{\boldsymbol{\Phi}}_{i-1} t_{i-1})(\mathbf{H}_{t,k}^T - \hat{\mathbf{H}}_{t,k}^T)\}$$
$$\{\alpha \frac{1}{N-1} (\mathbf{H}_{t,k} - \hat{\mathbf{H}}_{t,k})(\mathbf{H}_{t,k}^T - \hat{\mathbf{H}}_{t,k}^T) + (1-\alpha)\boldsymbol{\sigma}_t^T \boldsymbol{\sigma}_t\}^{-1}$$

$$\hat{\boldsymbol{\Phi}}_{t,k} = [\pi'_{t,k}(\varphi),...,\pi'_{t,k}(\varphi)], \qquad \pi'_{t,k}(\varphi) := \frac{1}{N}\sum_{j=1}^{N}\varphi_{t,k}^{(j)}, \qquad \mathbf{H}_{t,k} := [h_{t,k}^{(1)},...,h_{t,k}^{(N)}] \quad \text{and}$$

$$\hat{\mathbf{H}}_{t,k} = [\pi'_{t,k}(h),...,\pi'_{t,k}(h)].$$

Given below is a pseudo-code for the iterative updates.

### *Pseudo-code 2: for the iterative EnKS*

1. Follow steps 1:4 in pseudo-code 1.

   Set $k = 2$ and select $\beta_1$, $\kappa$.

2. (Iterative update)

   Using $\{\varphi_{i,k-1}^{(j)}\}_{j=1}^{N}$ compute $\{X_{i,k-1}^{(j)}\}_{j=1}^{N}$.

   Using $\{X_{i,k-1}^{(j)}\}_{j=1}^{N}$ compute $\{h_{i,k-1}^{(j)}\}_{j=1}^{N} = \{h(X_{i,k-1}^{(j)})\}_{j=1}^{N}$.

   Compute, $\qquad \boldsymbol{\Phi}_{i,k-1} := [\varphi_{i,k-1}^{(1)},...\varphi_{i,k-1}^{(N)}], \qquad \pi'_{i,k-1}(\varphi) := \frac{1}{N}\sum_{j=1}^{N}\varphi_{i,k-1}^{(j)}$

   $\hat{\boldsymbol{\Phi}}_{i,k-1} = [\pi'_{i,k-1}(\varphi),...,\pi'_{i,k-1}(\varphi)], \qquad \mathbf{H}_{i,k-1} := [h_{i,k-1}^{(1)},...,h_{i,k-1}^{(N)}], \quad \pi'_{i,k-1}(h) := \frac{1}{N}\sum_{j=1}^{N}h_{i,k-1}^{(j)},$

   $\hat{\mathbf{H}}_{i,k-1} = [\pi'_{i,k-1}(h),...,\pi'_{i,k-1}(h)].$

   Update each particle as

   $$\varphi_{i,k}^{(j)} = \varphi_{i,k-1}^{(j)} + \beta_{k-1}\mathbf{G}_{i,k-1}\{Y_i - h_{i,k-1}^{(j)}\}, \quad j = 1,...,N$$

   $$\mathbf{G}_{i,k-1} := \frac{1}{N}\begin{bmatrix}(\boldsymbol{\Phi}_{i,k-1}-\hat{\boldsymbol{\Phi}}_{i,k-1})(\mathbf{H}_{i,k-1}^{T}t_i - \hat{\mathbf{H}}_{i-1}^{T}t_{i-1} - \Delta\hat{\mathbf{H}}_{i,k-1}^{T}t_i)\\ +(\hat{\boldsymbol{\Phi}}_{i,k-1}t_i - \hat{\boldsymbol{\Phi}}_{i-1}t_{i-1})(\mathbf{H}_{i,k-1}^{T} - \hat{\mathbf{H}}_{i,k-1}^{T})\end{bmatrix}$$

   $$\left\{\alpha\frac{1}{N-1}(\mathbf{H}_{i,k-1}-\hat{\mathbf{H}}_{i,k-1})(\mathbf{H}_{i,k-1}^{T} - \hat{\mathbf{H}}_{i,k-1}^{T}) + (1-\alpha)\boldsymbol{\sigma}_i^{T}\boldsymbol{\sigma}_i\right\}^{-1}$$

3. Set $k = k+1$. If $k < \kappa$, set $\beta_{k+1} = \beta_k \exp(k+1-\kappa)$ and go to step 2;

else if $i < M$, go to step 3 in *pseudo-code 1* with $i = i+1$;

else terminate the algorithm.

Existence and uniqueness of solution (i.e. the conditional posterior distribution) via the proposed iterative scheme is demonstrated by the following two theorems (**Theorem 2 and 3**), whose proofs are given in Appendix I.

*Theorem 2*:

Suppose that,

a) $\boldsymbol{\varphi}_{i-1} \in L^2(P)$

b) $\boldsymbol{\varphi}_t \in C_b^2$

c) $\{\beta_0, \beta_1, ...\}$ is Cauchy sequence and converges to 1.

d) $L_t(\boldsymbol{\varphi})$ and $\boldsymbol{\sigma}_t$ are Lipschitz continuous uniformly on $(t_{i-1}, t_i]$.

Then there exists a limiting solution $\boldsymbol{\varphi}_t$ for Eqn. (5.1) as $k \to \infty$.

**Theorem 3:**

The solution via the proposed iterated algorithm is unique.

## 6. Numerical illustrations

### 6.1 Example 1: Dynamical system identification via a 200-dimensional nonlinear filtering problem

For the state-parameter estimation of a 50 degrees-of-freedom (DOFs) mechanical oscillator, the system model (herein correspondent to a 50-storied shear frame with uncertain damping and stiffness parameters) is considered to be of the form:

$$\ddot{U}(t) + [C]\dot{U}(t) + [K]U(t) = R(t) + f\dot{B}_t \tag{6.1}$$

$U$, $\dot{U} \in \mathbb{R}^{50}$ respectively denote the displacement and velocity vectors and scalar components define the corresponding quantities for different floors of the frame. The stiffness matrix is given as

$$[K] = \begin{bmatrix} K_1 + K_2 & -K_2 & 0 & ... & 0 \\ -K_2 & K_2 + K_3 & -K_3 & ... & 0 \\ 0 & -K_3 & K_3 + K_4 & ... & 0 \\ ... & ... & ... & ... & -K_{50} \\ 0 & 0 & 0 & -K_{50} & K_{50} \end{bmatrix}$$

The viscous damping matrix $[C]$ is similarly obtained by replacing $K_i$ by $C_i$ in the above matrix, where $K_i$ and $C_i$ are respectively the stiffness and damping parameters corresponding to the $i^{th}$ floor of the frame. $R(t) := \{r_i(t)\}_{i=1}^{n=50}$ is a random forcing vector with the transverse forcing at the $i$th floor given as $r_i(t) = 500\exp(-t)|\xi|\cos(5t)$ where $\xi \sim \mathcal{N}(0,1)$. The aim is to estimate the stiffness and damping coefficients as well as the velocity and displacement states, conditioned only on all the measured velocity components. Note that the present filtering problem is strictly nonlinear as the unknown parameters are taken as augmented states, i.e. the augmented state vector is given by [36; 27]:

$$X := \left\{ U^T; \dot{U}^T; \{K_1,...,K_{50}\}^T; \{C_1,...,C_{50}\}^T \right\} \in \mathbb{R}^{200}$$

Indeed the system process model would have been linear if the parameters were known. PFs are likely to diverge or collapse to a single particle for such a large dimensional state-parameter estimation problem with sparse data. Moreover, even for low dimensional problems, PFs may perform poorly with very low-intensity measurement noises, currently employed to reduce random fluctuations in the estimates due to large variance in the measurement noise. To demonstrate the performance of the proposed filter with low measurement noise levels (possible with sophisticated measuring devices), very low measurement noise intensity (less than 1%) is considered here for all the 50 components of the measured velocity vector. Since the EnKF (the ensemble Kalman filter as developed and implemented in [20]), is known to work for large dimensional filtering problems, it is used to report the numerical comparisons. An ensemble size $N = 800$ and time step $\Delta t = 0.01$ are taken for both the filters.

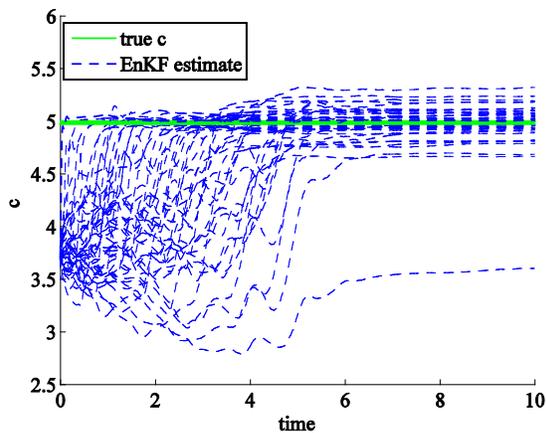
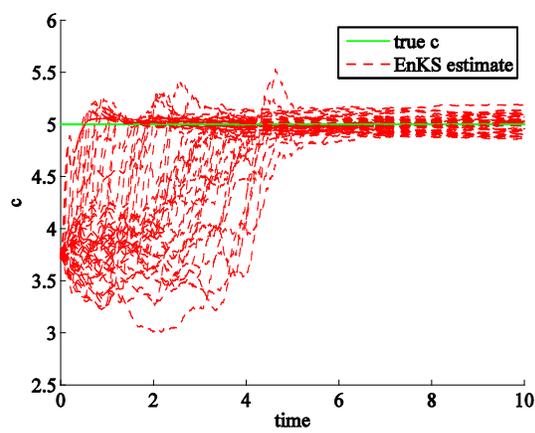

Figure 6.1 (a): Estimates of the damping parameters (C) by EnKF

Figure 6.1 (b): Estimates of the damping parameters (C) by EnKS

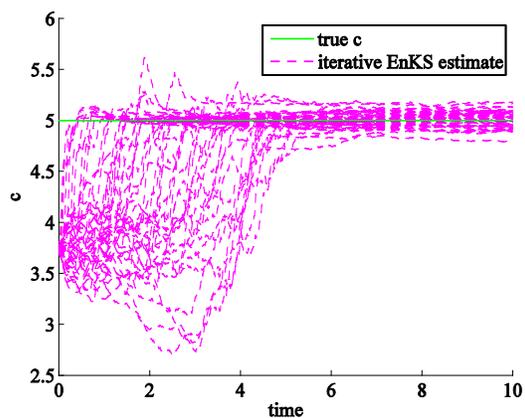

Figure 6.1 (c): Estimates of the damping parameters (C) by iterative EnKS

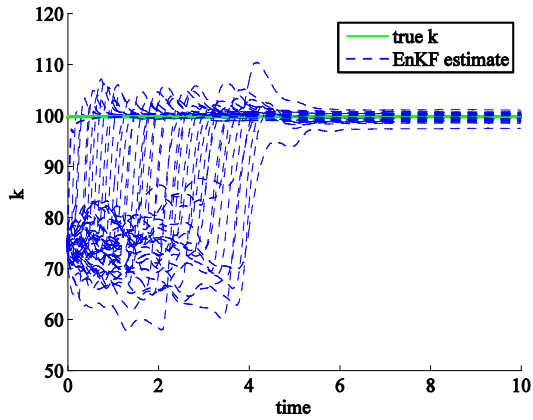
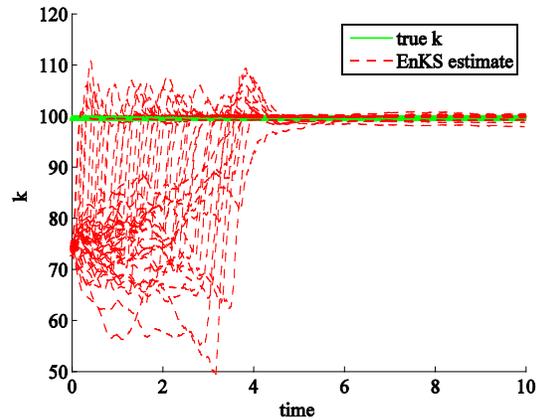

Figure 6.1 (d): Estimates of stiffness parameters (K) by EnKF

Figure 6.1 (e): Estimates of stiffness parameters (K) by EnKS

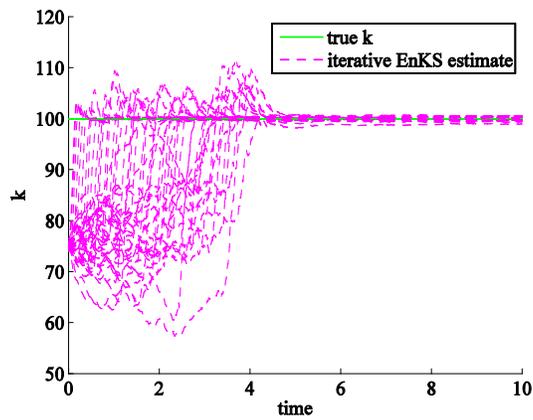

Figure 6.1 (f): Estimates of the stiffness parameters (K) by iterative EnKS

The reference stiffness parameters (used to integrate the system process *en route* to the generation of synthetic data by perturbing the computed solutions with appropriate noise) at each degree of freedom is taken as $k = 100$. Similarly the reference damping parameter at each floor is chosen as $c = 5$. It may be observed from Figure 6.1 that the EnKF underperforms in comparison to the EnKS and the iterative EnKS. Note that the simultaneous adoption of a large system dimension and low measurement noise ensures that the current identification problem is a difficult one.

## 6.2 Example 2: Damage detection for a 20-DOF mechanical oscillator

Consider once more the oscillator model for a shear frame, albeit of a lower dimension corresponding to 20 DOFs, so that the system model is formally given by:

$$\ddot{X}(t) + [C]\dot{X}(t) + [K]X(t) = R(t) + f\dot{B}_t \tag{6.2}$$

Thus the stiffness matrix is given as $[K] = \begin{bmatrix} K_1+K_2 & -K_2 & 0 & ... & 0 \\ -K_2 & K_2+K_3 & -K_3 & ... & 0 \\ 0 & -K_3 & K_3+K_4 & ... & 0 \\ ... & ... & ... & ... & ... \\ 0 & 0 & 0 & -K_{20} & K_{20} \end{bmatrix}$ and the viscous damping matrix $[C]$ is obtained by replacing $K_i$ by $C_i$ in the above matrix. As before, $R(t) := \{r_i(t)\}_{i=1}^{n=20}$ is a random forcing vector whose $i$-th element, $r_i(t) = 500\exp(-t)|\xi|\cos(5t)$ where $\xi \Box \mathcal{N}(0,1)$, denotes the transverse loading at the $i$-th storey. While the aim remains to estimate the stiffness and damping coefficients along with the velocity/displacement states (conditioned on only the measured velocities of the floors), the reference stiffness parameter $K_{10}$ is kept at a slightly lower value vis-à-vis the rest $\{K_m \mid m \in [1,20]\} \setminus \{K_{10}\}$. This is a simple representation of a slightly reduced load-carrying capacity at the $10^{th}$ floor, thereby indicating incipient damage/degradation. Specifically, we take $K_{10} = 98$ even as the rest of the stiffness parameters are maintained at 100. Damping parameter at each floor is chosen as 5. Here the augmented state vector is 80-dimensional and, as in the last example, a low noise intensity ($< 1\%$) is applied in generating the data. An ensemble size of $N = 300$ and time step $\Delta t = 0.01$ are taken for both the EnKF and EnKS filter runs.

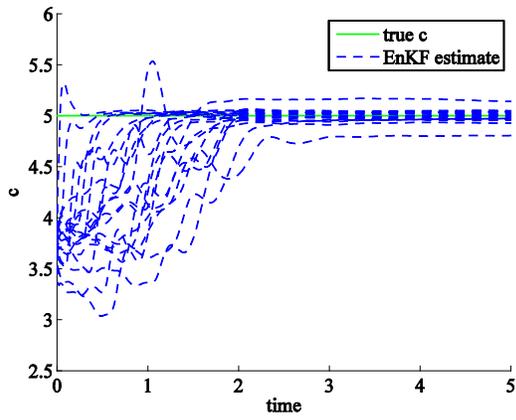
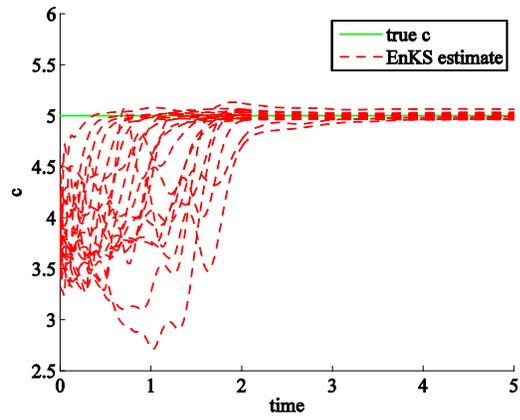

Figure 6.2 (a): Estimates of the damping parameters (C) by EnKF

Figure 6.2 (b): Estimates of the damping parameters (C) by EnKS

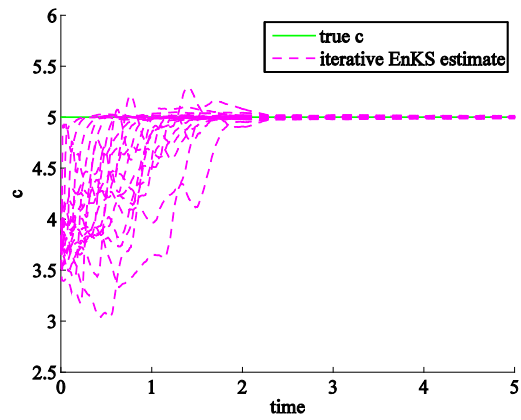

Figure 6.2 (c): Estimates of the damping parameters (C) by iterative EnKS

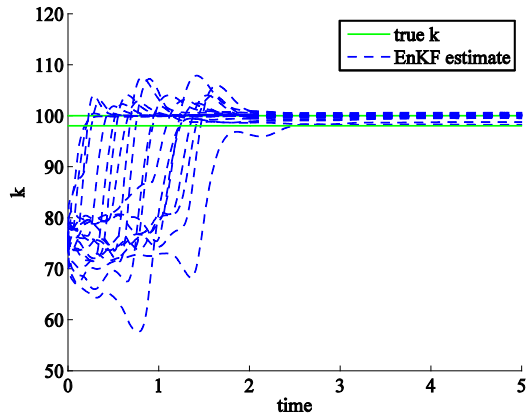
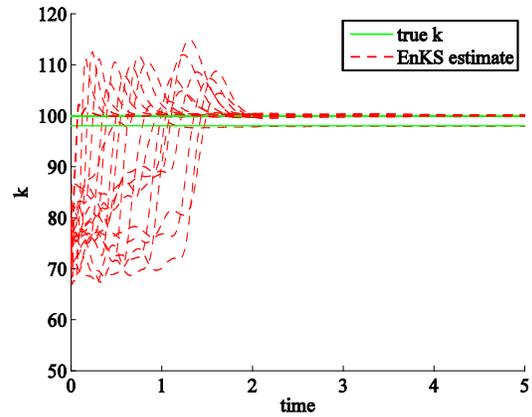

Figure 6.2 (d): Estimates of the stiffness parameters (K) by EnKF

Figure 6.2 (e): Estimates of the stiffness parameters (K) by EnKS

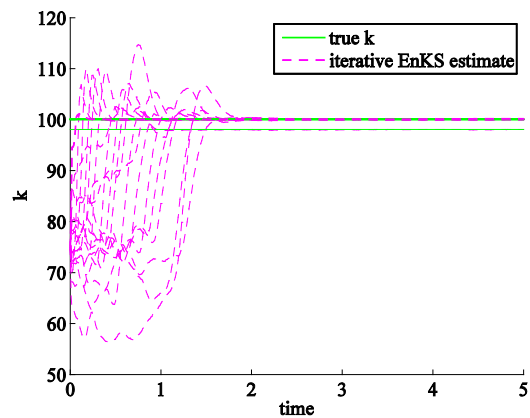

Figure 6.2 (f): Estimates of the stiffness parameters (K) by iterative EnKS

A relatively poorer performance of the EnKF in comparison with the EnKS is evident from the estimation results plotted in Figure 6.2. EnKF not only fails to detect the incipient damage (Figures 6.2d, 6.2e), but also yields the estimation of the damping coefficients with poorer resolution (Figures 6.2a – 6.2c). Moreover, as anticipated, the iterative EnKS (broken magenta), performs better than the non-iterative EnKS (broken red) even though the contrast in performance between the two variants is never quite striking. Hence, in the last two examples to follow, only the non-iterative version of the proposed filter is made use of.

### 6.3 Example 3: Identifying a nonlinear oscillator with nonlinear measurement

Strictly speaking, the EnKF [20] is not quite suited to treating nonlinearity in the measurement model. However, in practice, variants of the EnKF have indeed been applied to filtering problems involving nonlinear measurement models. One thus anticipates a sharper performance contrast between the EnKF and the EnKS in such cases, even if the dimension of the filtering problem were smaller. This point is currently emphasized through a problem of estimating state and parameters of a 1-DOF nonlinear oscillator model, wherein a transducer supplied at the base measures the reaction transferred there (Figure 6.3a).

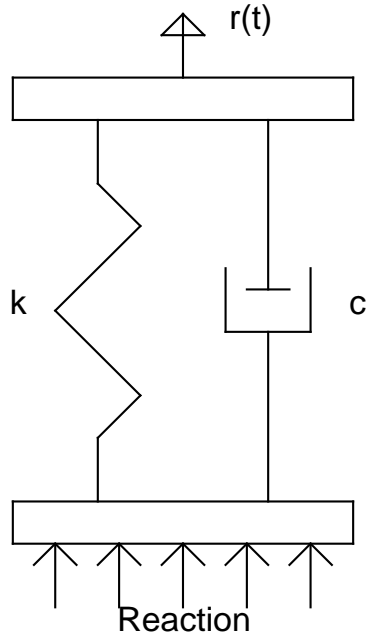

Figure 6.3 (a): A nonlinear oscillator; reaction transferred at the base is measured

The system process and measurement models are respectively given as:

$$\ddot{X}_t + c\dot{X}_t + k\sin(X_t) = r(t) + f\dot{B}_t \tag{6.3}$$

$$y_t = c\dot{X}_t + k\sin(X_t) + \Delta\eta_t \tag{6.4}$$

Displacement ($X_t$), velocity ($\dot{X}_t$), damping (*c*) and stiffness (*k*) are estimated using an ensemble size $N = 600$ and time step $\Delta t = 0.01$. A random force $r(t) := 5\exp(-0.01t)|\xi|\cos(5t)$, $\xi \sim \mathcal{N}(0,1)$, is applied at the free end of the oscillator. An assessment of the estimates, reproduced in Figures 6.3b-e, readily reveals a degraded EnKF performance (broken blue), especially in estimating the viscous damping parameter.

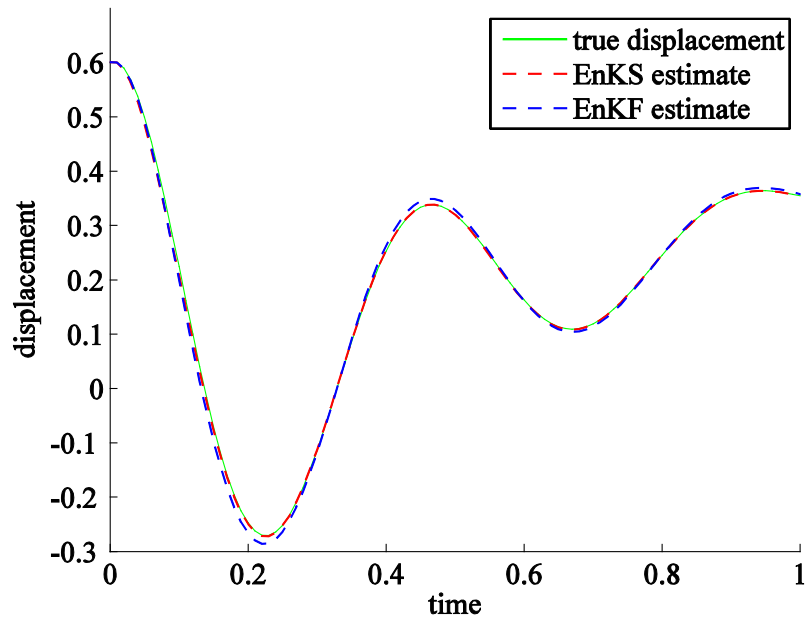

Figure 6.3 (b): Estimate of displacement

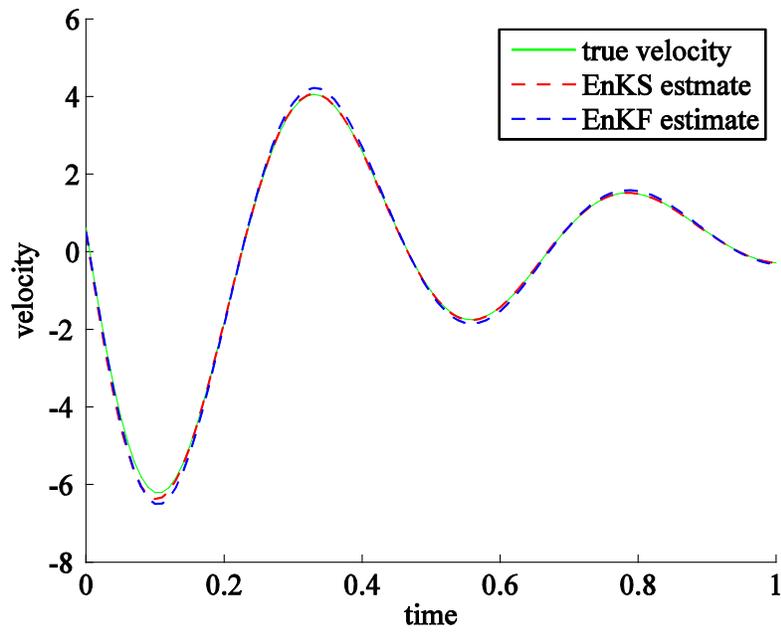

Figure 6.3 (c): Estimate of velocity

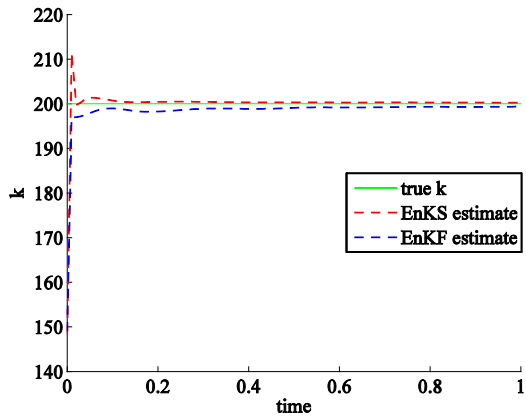
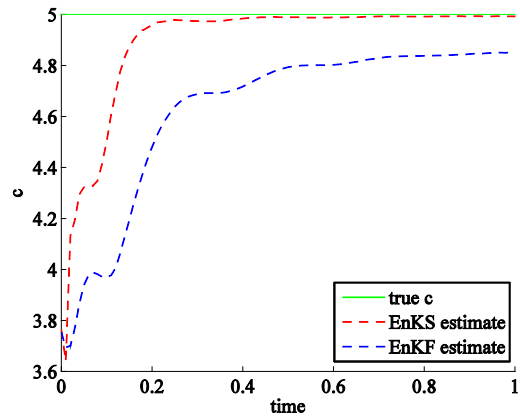

Figure 6.3 (d): Estimate of $k$        Figure 6.3 (e): Estimate of $c$

### 6.4 Example 4: A nonlinear population model

As the final example, we consider a nonlinear population equation [37] whose governing dynamics is given below:

$$\dot{X}_t = -r_1\left(1 - \frac{X_t}{r_2}\right)X_t, \qquad X_{t_0} = X_0 \tag{6.5}$$

where $r_1 = 1$, $r_2 = 2 > 0$. One important property of Eqn. (6.5) is its sensitivity to the initial condition $X_0$, i.e., if $X_0 > r_2$ the solution diverges exponentially, otherwise it approaches 0. Following [37], we create a reference (true) state, obtainable as the integrated trajectory of $X_t$ (with $X_0 = 2.1$) added to by a Brownian noise $\mathcal{N}(0, 0.2^2 t)$. Measurements are sampled at every $\Delta t = 0.1$ with the measurement noise characterized by a Gaussian distribution $\mathcal{N}(0, 0.1^2)$. The aim is to produce the filtered state that follows the trends of the true state with some fidelity, even as the individually integrated trajectories of the system process dynamics may tend to diverge quickly. Using an ensemble size $N = 1000$, comparisons of the estimates are reported in Figure 6.4 where the broken blue line corresponds to the EnKF estimate and the broken red line to the EnKS. The diverging trend in the EnKS reconstruction is significantly more muted in comparison with the EnKF. The present exercise may also be contrasted with the significantly higher ensemble size $N = 10^6$ used in [37] to report the performance of an ensemble square root filter, a variant of the EnKF.

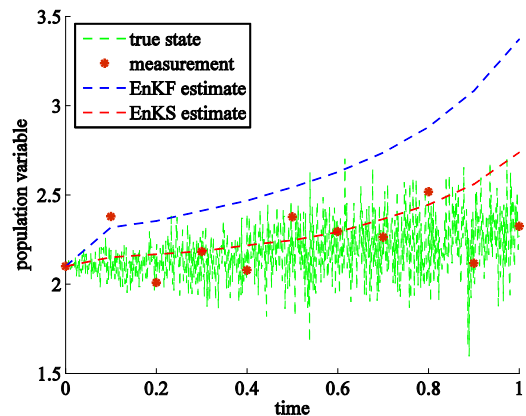

Figure 6.4: State estimation for the nonlinear population equation

## 7. Conclusions

The nonlinear and additive particle update in the proposed EnKS, derived through consistent ensemble and time discretizations of the Kushner-Stratonovich SPDE, is essentially aimed at alleviating some of the prominent numerical bottlenecks characteristic of weight-based updates in particle filters. An efficient implementation of the update is made possible through several manipulations on the discretized innovation integral designed to efficiently drive the measurement-prediction misfit to a zero-mean martingale, herein characterized by an Ito integral. Arguably, the most notable development of this work is the non-iterative version of the EnKS that is shown to work quite accurately for nonlinear filtering problems of large dimension and involving sparse data with possibly low measurement noise intensity. This is the regime where a particle filter typically fails to perform. Motivated by the stochastic Picard iteration and implemented using annealing-type inner iterates, an iterative version of the EnKS is also synthesized which, though computationally more intensive, is able to achieve still higher accuracy in the computed estimates. A reflection of the nonlinear nature of the update is in the demonstrably superior performance of both variants of the EnKS over the well known ensemble Kalman filter.

The structure of the EnKS, which shares the familiar gain-based update features of the popular Kalman filter, makes it ideal for nonlinear filtering applications with feedback control. Finally, consistent with the martingale problem setting as originated by Stroock and Varadhan [40], the EnKS admits a far more general class of non-smooth drift and diffusion fields in the system process and measurement models than is permissible with filters requiring linearizations of these terms.

## References

1. Kalman, R. E. and Bucy, R. S. (1961) New results in linear filtering and prediction theory. *Trans. ASME, Ser. D, J. Basic Eng.* 83, 95–107.

2. Kalman, R.E. (1960) A new approach to linear filtering and prediction problems. *Transactions of the ASME- Journal of Basic Engineering*, 82 (1), 35–45.

3. Hoshiya, M and Saito, E. (1984) Structural identification by extended Kalman filter, *Journal of Engineering Mechanics*, 110(12), 1757-1770.

4. Budhiraja, A., Chen, L. and Lee, C. (2007) A survey of numerical methods for nonlinear filtering problems. *Physica D* 230, 27-36.

5. Kallianpur, G. and Striebel, C. (1968) Estimation of stochastic systems: Arbitrary system process with additive white noise observation errors. *The Annals of Mathematical Statistics*, *39*(3), 785-801.

6. Zakai, M. (1969) On the optimal filtering of diffusion processes, *Zeitschrift für Wahrscheinlichkeitstheorie und verwandte Gebiete*, 11(3), 230-243.

7. Kushner, (1964) H. J. On the differential equations satisfied by conditional probablitity densities of Markov processes, with applications. *SIAM J. Series A: Control* 2.1: 106-119.

8. Ito, K. and Rozovskii, B. (2000) Approximation of the Kushner equation for nonlinear filtering, *SIAM J. Control Optim.* 38(3):893-915.

9. Julier, S. J. and Uhlmann, J. K. (1997) A New Extension of the Kalman Filter to Nonlinear Systems, *In Proc. of AeroSense: The 11th Int. Symp. On Aerospace/Defence Sensing, Simulation and Controls*.

10. Doucet, A., Godsill, S. and Andrieu, C. (2000) On sequential Monte Carlo sampling methods for Bayesian filtering, *Statistics and Computing* 10, 197–208.

11. Fearnhead, P., Papaspiliopoulos, O. and Roberts, G. O. (2008) Particle filters for partially observed diffusions, *Journal of the Royal Statistical Society: Series B (Statistical Methodology)*, 70(4), 755-777.

12. Gordon, N. J., Salmond, D. J. and Smith, A. F. M. (1993) Novel approach to nonlinear/non-Gaussian Bayesian state estimation, *Radar and Signal Processing, IEE Proceedings F,* 140, 107-113.

13. Pitt, M. and Shephard, N. (1999) Filtering via simulation: Auxiliary particle filters, *J. Amer. Statist. Assoc.* 94, 590–599.

14. Budhiraja, A., Chen, L. and Lee, C. (2007) A survey of numerical methods for nonlinear filtering problems, *Physica D* 230, 27-36.

15. Crisan, D., & Doucet, A. (2000) Convergence of sequential Monte Carlo methods, *Signal Processing Group, Department of Engineering, University of Cambridge, Technical Report CUEDIF-INFENGrrR38, 1*.

16. Chopin, N. (2004) Central limit theorem for sequential Monte Carlo methods and its application to Bayesian inference *The Annals of Statistics* 32(6), 2385-2411.

17. Chorin, A. J. and Tu, X. (2009) Implicit sampling for particle filters, *Proc. Natl. Acad. Sci. USA* 106: 17249-17254.

18. Jeroen, D. H., Thomas, B.S., Gustafsson, F. (2006) On Resampling Algorithms For Particle Filters, *IEEE Nonlinear Statistical Signal Processing Workshop* 79-82, Cambridge, UK.

19. Geweke, J. and Tanizaki, H. (1999) On Markov Chain Monte Carlo methods for nonlinear and non-gaussian state-space models, *Commun. Stat. Simul. C* 28, 867–894.

20. Evensen, G. (2009). *Data assimilation: the ensemble Kalman filter*, Springer.

21. Budhiraja, A. and Kallianpur, G. (1996) Approximations to the solutions of the Zakai equation using multiple Wiener and Stratonovich integral expansions, *Stochastics and Stochastic Reports* 56, 271-315.

22. Gobet, E., Pagès,G., Pham, H. and Printems, J. (2006) Discretization and simulation of the Zakai equation, *SIAM J. Numer. Anal.* 44(6), 2505-2538.

23. Ahmed, N. U. and Radaideh, S. M. (1997) A powerful numerical technique solving Zakai equation for nonlinear filtering, *Dynamics and Control* 7, 293-308.

24. Budhiraja, A. and Kallianpur, G. (1997) The Feynman-Stratonovich Semigroup and Stratonovich Integral Expansions in Nonlinear Filtering, *Appl. Math. Optim.* 35, 91-116.

25. Milstein, G. V. and Tretyakov, M.V. (2009) Solving parabolic stochastic partial differential equations via averaging over characteristics, *Math. Comp.* 78, 2075–2106.

26. Crisan, D. and Lyons, T. (1999) A particle approximation of the solution of the Kushner-Stratonovich equation, *Probab. Theory Related Fields* 115, 549-578.

27. Sarkar, S , Chowdhury, S. R., Venugopal, M., Vasu, R. M., & Roy, D. (2014) A Kushner–Stratonovich Monte Carlo filter applied to nonlinear dynamical system identification, *Physica D: Nonlinear Phenomena* 270, 46-59.

28. Oksendal, B.K. (2003) *Stochastic Differential Equations-An Introduction with Applications*, 6th ed., Springer, New York.

29. Kurtz, T. G. (1998) Martingale problems for conditional distributions of Markov processes, *Electron. J. Probab* 3(9), 1-29.

30. Milstein, G. N., Platen, E., & Schurz, H. (1998) Balanced implicit methods for stiff stochastic systems, *SIAM Journal on Numerical Analysis*, 35(3), 1010-1019.

31. Hartman, P. (1963) On the local linearization of differential equations, *Proceedings of the American Mathematical Society*, 14(4), 568-573.

32. Roy, D. (2004) A family of lower- and higher-order transversal linearization techniques in non-linear stochastic engineering dynamics, *Int. J. Numer. Meth. Eng.* 61:764–790.

33. Roy, D., Saha, N. and Dash, M. K. (2008) Weak Forms of the Locally Transversal Linearization (LTL) Technique for Stochastically Driven Nonlinear Oscillators, *Appl. Math. Model.* 32:1657-1681.

34. Roy, D., & Dash, M. K. (2002) A stochastic Newmark method for engineering dynamical systems, *Journal of sound and vibration*, 249(1), 83-100.

35. Kirkpatrick, S., Gelatt, C. D. and Vecchi, M. P. (1983) Optimization by simulated annealing, *Sci.* 220:671-680.

36. Raveendran, T., Sarkar, S., Roy, D., & Vasu, R. M. (2013) A novel filtering framework through Girsanov correction for the identification of nonlinear dynamical systems. *Inverse Problems*, 29(6), 065002.

37. Li, J., & Xiu, D. (2009) A generalized polynomial chaos based ensemble Kalman filter with high accuracy, *Journal of computational physics*, 228(15), 5454-5469.

38. Kloeden, P. and Platen, E. (1992) *Numerical Solutions of Stochastic Differential Equations,* Springer, Berlin.

39. van Handel, R. (2007) Stochastic Calculus, Filtering, and Stochastic Control.Course notes., *URL http://www. princeton. edu/~ rvan/acm217/ACM217. pdf*.

40. Stroock, D. W. and Varadhan, S.R. (1972) On the support of diffusion processes with application to the strong maximum principle, *In Proceedings of the Sixth Bercley Symposium on Mathematical Statistics and Probability (Univ. California, Bercley, Calif. 1970/1971)*, 3, 330-359.

**Appendix I:**

In the least square sense, the error in the filtered estimate due to approximations in time and finiteness of the ensemble may be denoted by $\left(E_P\left[\left|\pi_i(\varphi)-\pi_i'^e(\varphi)\right|^2\right]\right)^{\frac{1}{2}}$, where $(\cdot)^e$ and $(\cdot)'$ denote EM and ensemble approximations respectively. As a precursor to getting an error bound, this term is split into two parts, one corresponding to EM integration error and the other due to ensemble approximation:

$$\left(E_P\left[\left|\pi_i(\varphi)-\pi_i'^e(\varphi)\right|^2\right]\right)^{\frac{1}{2}} \leq \underbrace{\left(E_P\left[\left|\pi_i(\varphi)-\pi_i^e(\varphi)\right|^2\right]\right)^{\frac{1}{2}}}_{(\text{term-1: time discretization error})} + \underbrace{\left(E_P\left[\left|\pi_i^e(\varphi)-\pi_i'^e(\varphi)\right|^2\right]\right)^{\frac{1}{2}}}_{(\text{term-2: ensemble approximation error})}$$

(A1)

We first obtain an error bound for the time discretization.

*Lemma 1:*

If the process has bounded moments of any order and $\varphi \in C_b^2(\mathbb{R})$, then:

$$\left(E_P\left[\left|\pi_i(\varphi)-\pi_i^e(\varphi)\right|^2\right]\right)^{\frac{1}{2}} \leq D'(\Delta t_i)^{\frac{1}{2}}$$

*Proof:*

Using the conditional version of Jensen's inequality, we have:

$$E_P\left[\left|\pi_i(\varphi)-\pi_i^e(\varphi)\right|^2\right] = E_P\left[\left|E_P\left[\varphi_i-\varphi_i^e\right]|\mathcal{F}_i^Y\right|^2\right]$$

$$\leq E_P\left[E_P\left[\left|\varphi_i-\varphi_i^e\right|^2 | \mathcal{F}_i^Y\right]\right]$$

$$= E_P\left[\left|\varphi_i-\varphi_i^e\right|^2\right]$$

Using the standard strong order of convergence of the EM method [38], we obtain:

$$\left(E_P\left[\left|X_i - X_i^e\right|^2\right]\right)^{\frac{1}{2}} \leq D_1 \left(\Delta t_i\right)^{\frac{1}{2}} \tag{A2}$$

where $D_1 > 0$ is a constant independent of $\Delta t_i$. In general, for $p \geq 1, D_2 > 0$, one can write:

$$E_P\left[\left|X_i - X_i^e\right|^{2p}\right]^{\frac{1}{2p}} \leq D_2 \left(\Delta t_i\right)^{\frac{1}{2}} \tag{A3}$$

Furthermore, we assume that $\varphi$ is sufficiently smooth so that $\varphi(x)$ and its derivatives satisfy an inequality of the form:

$$\left|\varphi(x)\right| \leq D_3 \left(1 + |x|^a\right) \tag{A4}$$

for some constants $D_3, a > 0$. Hence we can write:

$$\left|\varphi_i - \varphi_i^e\right| \leq D_4 \left(1 + |X_i|^a + |X_i^e|^a\right)\left|X_i - X_i^e\right| \tag{A5}$$

where $D_4 > 0$. Then, using the Cauchy-Schwarz inequality, we have

$$\begin{aligned}
E_P\left[\left|\varphi_i - \varphi_i^e\right|^{2p}\right] &\leq \left(D_4\right)^{2p} E_P\left[\left(1 + |X_i|^a + |X_i^e|^a\right)^{2p} \left|X_i - X_i^e\right|^{2p}\right] \\
&\leq \left(D_4\right)^{2p} \sqrt{E_P\left[\left(1 + |X_i|^a + |X_i^e|^a\right)^{4p}\right]} \sqrt{E_P\left[\left|X_i - X_i^e\right|^{4p}\right]} \\
&\leq D'(X_i)\left(\left(\Delta t_i\right)^{\frac{1}{2}}\right)^{2p}
\end{aligned} \tag{A6}$$

Hence, for $p = 1$, we get $\left(E_P\left[\left|\pi_i(\varphi) - \pi_i^e(\varphi)\right|^2\right]\right)^{\frac{1}{2}} \leq D'\left(\Delta t_i\right)^{\frac{1}{2}}$ (A7)

where $D'(X_i) > 0$ is independent of $\Delta t_i$.

□

Next, we consider the error due to the ensemble approximation within a time-discretized framework. In a recursive setting, given the empirical filtered distribution of $X_t$ at $t = t_{i-1}$, we may consider:

$$E_P\left[\left|\pi_{i-1}^e(\varphi) - \pi_{i-1}^{'e}(\varphi)\right|^2\right] \leq D_5 \frac{\|\varphi\|^2}{N}$$

where $\|\cdot\|$ denotes the supremum norm on $C_b(\mathbb{R}^n)$.

*Lemma 2:*

Assume that for any $\varphi \in C_b^2(\mathbb{R}^n)$,

$$E_P\left[\left|\pi_{i-1}^e(\varphi) - \pi_{i-1}^{'e}(\varphi)\right|^2\right] \leq D_5 \frac{\|\varphi\|^2}{N}$$

Then, for $D_5 > 0$

$$\left(E_P\left[\left|\pi_i^{'e}(\varphi) - \pi_i^e(\varphi)\right|^2\right]\right)^{\frac{1}{2}}$$

$$\leq \frac{D_{13}}{\sqrt{N}} \Delta t_i \left(|Y_i| + \|h_i\|\right) \left\{ \begin{array}{l} \left( \begin{array}{l} \left\|L\left((\varphi - \overline{\varphi})(ht - \overline{h}t + \overline{h}_{i-1}\Delta t_i)^T\right)\right\| \\ + \|\varphi\|\|h\| + \left\|L\left((\overline{\varphi}t - \overline{\varphi}_{i-1}t_{i-1})(h - \overline{h})^T\right)\right\| \end{array} \right) \left\|\left\{\begin{array}{l}\alpha(h-\overline{h})^T(h-\overline{h})\\+(1-\alpha)\sigma_i^T\sigma_i\end{array}\right\}^{-1}\right\| \\ \\ + \|\varphi\|\|h\|\alpha\left(\Delta t_i \left\|L\left((h-\overline{h})(h-\overline{h})^T\right)\right\| + \left\|(h-\overline{h})(h-\overline{h})^T\right\|\right) \left\|\left\{\begin{array}{l}\alpha(h-\overline{h})^T(h-\overline{h})\\+(1-\alpha)\sigma_i^T\sigma_i\end{array}\right\}^{-1}\right\|^2 \end{array} \right\}$$

$$+ \frac{D_{13}}{\sqrt{N}}\left(\left\|\tilde{G}_i(\varphi)\right\|\left(\Delta t_i \|L(h)\| + \|h\|\right) + \left(\Delta t_i \|L(\varphi)\| + \|\varphi\|\right)\right)$$

*Proof:*

Using Minkowski's inequality, we can write,

$$\left(E_P\left[\left|\pi_i^{\prime e}(\boldsymbol{\varphi})-\pi_i^e(\boldsymbol{\varphi})\right|^2\right]\right)^{\frac{1}{2}} \leq \left(E_P\left[\left|\{\tilde{\pi}_i^{\prime e}(\boldsymbol{\varphi})+\tilde{\mathbf{G}}_i^{\prime}(\boldsymbol{\varphi})\tilde{I}_i^{\prime}(\boldsymbol{\varphi})\}-\{\tilde{\pi}_i^{\prime e}(\boldsymbol{\varphi})+\tilde{\mathbf{G}}_i(\boldsymbol{\varphi})\tilde{I}_i(\boldsymbol{\varphi})\}\right|^2\right]\right)^{\frac{1}{2}}$$

$$+\left(E_P\left[\left|\{\tilde{\pi}_i^{\prime e}(\boldsymbol{\varphi})+\tilde{\mathbf{G}}_i(\boldsymbol{\varphi})\tilde{I}_i(\boldsymbol{\varphi})\}-\{\tilde{\pi}_i^e(\boldsymbol{\varphi})+\tilde{\mathbf{G}}_i(\boldsymbol{\varphi})\tilde{I}_i(\boldsymbol{\varphi})\}\right|^2\right]\right)^{\frac{1}{2}}$$

$$=\left(E_P\left[\left|\tilde{\mathbf{G}}_i^{\prime}(\boldsymbol{\varphi})\tilde{I}_i^{\prime}(\boldsymbol{\varphi})-\tilde{\mathbf{G}}_i(\boldsymbol{\varphi})\tilde{I}_i(\boldsymbol{\varphi})\right|^2\right]\right)^{\frac{1}{2}}+\left(E_P\left[\left|\tilde{\pi}_i^{\prime e}(\boldsymbol{\varphi})-\tilde{\pi}_i^e(\boldsymbol{\varphi})\right|^2\right]\right)^{\frac{1}{2}}$$

(A8)

where $\tilde{\pi}_i^e(\boldsymbol{\varphi}):=\pi_i^e(\tilde{\boldsymbol{\varphi}})$, $\tilde{I}_i^{\prime}(\boldsymbol{\varphi}):=Y_i-\tilde{\pi}_i^{\prime e}(h)$ and

$$\tilde{\mathbf{G}}_i^{\prime}=\frac{1}{N}\left\{\left(\tilde{\boldsymbol{\Phi}}_i-\hat{\tilde{\boldsymbol{\Phi}}}_i\right)\left(\tilde{\mathbf{H}}_i^T t_i-\hat{\tilde{\mathbf{H}}}_{i-1}^T t_{i-1}-\Delta\hat{\tilde{\mathbf{H}}}_i^T t_i\right)+\left(\hat{\tilde{\boldsymbol{\Phi}}}_i t_i-\hat{\tilde{\boldsymbol{\Phi}}}_{i-1}t_{i-1}\right)\left(\tilde{\mathbf{H}}_i^T-\hat{\tilde{\mathbf{H}}}_i^T\right)\right\}$$

$$\left\{\alpha\frac{1}{N-1}\left(\tilde{\mathbf{H}}_i-\hat{\tilde{\mathbf{H}}}_i\right)\left(\tilde{\mathbf{H}}_i^T-\hat{\tilde{\mathbf{H}}}_i^T\right)+(1-\alpha)\boldsymbol{\sigma}_i^T\boldsymbol{\sigma}_i\right\}^{-1}$$

$$=\left\{\tilde{\pi}_i^{\prime e}\left((\boldsymbol{\varphi}-\overline{\boldsymbol{\varphi}})\left(ht-\overline{h}t+\overline{h}_{i-1}\Delta t_i\right)^T\right)+\tilde{\pi}_i^{\prime e}\left((\overline{\boldsymbol{\varphi}}t-\overline{\boldsymbol{\varphi}}_{i-1}t_{i-1})\left(h-\overline{h}\right)^T\right)\right\}\left\{\alpha\tilde{\pi}_i^{\prime e}\left((h-\overline{h})(h-\overline{h})^T\right)+(1-\alpha)\boldsymbol{\sigma}_i^T\boldsymbol{\sigma}_i\right\}^{-1}$$

As before, we sometimes replace the conditional expectation of a variable with an over-bar for conciseness. One may write similar expressions for $\tilde{\mathbf{G}}_i$ and $\tilde{I}_i$ by appropriately replacing the ensemble approximation operator in $\tilde{\mathbf{G}}_i^{\prime}$ and $\tilde{I}_i^{\prime}$.

Using Minkowski's inequality on the second term of the RHS of Eqn. (A8), we have

$$\left(E_P\left[\left|\tilde{\pi}_i^{\prime e}(\boldsymbol{\varphi})-\tilde{\pi}_i^e(\boldsymbol{\varphi})\right|^2\right]\right)^{\frac{1}{2}} \leq \left(E_P\left[\left|\tilde{\pi}_i^{\prime e}(\boldsymbol{\varphi})-\left(\pi_{i-1}^{\prime e}(\boldsymbol{\varphi})+\pi_{i-1}^e(\mathrm{L}(\boldsymbol{\varphi}))\Delta t_i\right)\right|^2\right]\right)^{\frac{1}{2}}$$

$$+\left(E_P\left[\left|\left(\pi_{i-1}^{\prime e}(\boldsymbol{\varphi})+\pi_{i-1}^e(\mathrm{L}(\boldsymbol{\varphi}))\Delta t_i\right)-\tilde{\pi}_i^e(\boldsymbol{\varphi})\right|^2\right]\right)^{\frac{1}{2}}$$

(A9)

$$\leq \Delta t_i\left(E_P\left[\left|\pi_{i-1}^{\prime e}(\mathrm{L}(\boldsymbol{\varphi}))-\pi_{i-1}^e(\mathrm{L}(\boldsymbol{\varphi}))\right|^2\right]\right)^{\frac{1}{2}}+\left(E_P\left[\left|\pi_{i-1}^{\prime e}(\boldsymbol{\varphi})-\pi_{i-1}^e(\boldsymbol{\varphi})\right|^2\right]\right)^{\frac{1}{2}}$$

$$\leq D_6(\Delta t_i\frac{\|\mathrm{L}(\boldsymbol{\varphi})\|}{\sqrt{N}}+\frac{\|\boldsymbol{\varphi}\|}{\sqrt{N}})$$

Towards getting a bound for the first term on the RHS of Eqn. (A8), the term is split as:

$$\left(E_P\left[\left|\tilde{\mathbf{G}}'_i(\boldsymbol{\varphi})\tilde{I}'_i(\boldsymbol{\varphi}) - \tilde{\mathbf{G}}_i(\boldsymbol{\varphi})\tilde{I}_i(\boldsymbol{\varphi})\right|^2\right]\right)^{\frac{1}{2}} \leq \left(E_P\left[\left|\tilde{\mathbf{G}}'_i(\boldsymbol{\varphi})\tilde{I}'_i(\boldsymbol{\varphi}) - \tilde{\mathbf{G}}_i(\boldsymbol{\varphi})\tilde{I}'_i(\boldsymbol{\varphi})\right|^2\right]\right)^{\frac{1}{2}}$$

$$+ \left(E_P\left[\left|\tilde{\mathbf{G}}_i(\boldsymbol{\varphi})\tilde{I}'_i(\boldsymbol{\varphi}) - \tilde{\mathbf{G}}_i(\boldsymbol{\varphi})\tilde{I}_i(\boldsymbol{\varphi})\right|^2\right]\right)^{\frac{1}{2}} \quad \text{(A10)}$$

$$\leq (|Y_i| + \|h_i\|)\left(E_P\left[\left|\tilde{\mathbf{G}}'_i(\boldsymbol{\varphi}) - \tilde{\mathbf{G}}_i(\boldsymbol{\varphi})\right|^2\right]\right)^{\frac{1}{2}}$$

$$+ \|\tilde{\mathbf{G}}_i(\boldsymbol{\varphi})\|\left(E_P\left[\left|\tilde{\pi}^e_i(h) - \tilde{\pi}'^e_i(h)\right|^2\right]\right)^{\frac{1}{2}}$$

Considering the last factor of the first term on the RHS of Eqn. (A10), we have:

$$\left(E_P\left[\left|\tilde{\mathbf{G}}'_i(\boldsymbol{\varphi}) - \tilde{\mathbf{G}}_i(\boldsymbol{\varphi})\right|^2\right]\right)^{\frac{1}{2}}$$

$$\leq \left(E_P\left[\left|\begin{Bmatrix}\tilde{\pi}'^e_i\left((\boldsymbol{\varphi}-\overline{\boldsymbol{\varphi}})(ht - \overline{h}t + \overline{h}_{i-1}\Delta t_i)^T\right) \\ +\tilde{\pi}'^e_i\left((\overline{\boldsymbol{\varphi}}t - \overline{\boldsymbol{\varphi}}_{i-1}t_{i-1})(h - \overline{h})^T\right)\end{Bmatrix} - \begin{Bmatrix}\tilde{\pi}^e_i\left((\boldsymbol{\varphi}-\overline{\boldsymbol{\varphi}})(ht - \overline{h}t + \overline{h}_{i-1}\Delta t_i)^T\right) \\ +\tilde{\pi}^e_i\left((\overline{\boldsymbol{\varphi}}t - \overline{\boldsymbol{\varphi}}_{i-1}t_{i-1})(h - \overline{h})^T\right)\end{Bmatrix}\right|^2\right]\right)^{\frac{1}{2}} \left(E_P\left[\left|\left\{\alpha\tilde{\pi}'^e_i\left((h-\overline{h})(h-\overline{h})^T\right) + (1-\alpha)\boldsymbol{\sigma}^T_i\boldsymbol{\sigma}_i\right\}^{-1}\right|^2\right]\right)^{\frac{1}{2}}$$

$$+ \left(E_P\left[\left|\begin{Bmatrix}\tilde{\pi}^e_i\left((\boldsymbol{\varphi}-\overline{\boldsymbol{\varphi}})(ht - \overline{h}t + \overline{h}_{i-1}\Delta t_i)^T\right) \\ +\tilde{\pi}^e_i\left((\overline{\boldsymbol{\varphi}}t - \overline{\boldsymbol{\varphi}}_{i-1}t_{i-1})(h - \overline{h})^T\right)\end{Bmatrix}\right|^2\right]\right)^{\frac{1}{2}} \left(E_P\left[\left|\begin{Bmatrix}\alpha\tilde{\pi}'^e_i\left((h-\overline{h})(h-\overline{h})^T\right) + (1-\alpha)\boldsymbol{\sigma}^T_i\boldsymbol{\sigma}_i\end{Bmatrix}^{-1} \\ -\begin{Bmatrix}\alpha\tilde{\pi}^e_i\left((h-\overline{h})(h-\overline{h})^T\right) + (1-\alpha)\boldsymbol{\sigma}^T_i\boldsymbol{\sigma}_i\end{Bmatrix}^{-1}\right|^2\right]\right)^{\frac{1}{2}}$$

(A11)

Considering the first term on the RHS of the inequality A11, we get:

$$\left( E_P \left[ \left| \begin{aligned} &\left\{ \tilde{\pi}_i^{\prime e} \left( (\boldsymbol{\varphi}-\overline{\boldsymbol{\varphi}})\left(ht-\overline{h}t+\overline{h}_{i-1}\Delta t_i\right)^T \right) + \tilde{\pi}_i^{\prime e} \left( (\overline{\boldsymbol{\varphi}}t-\overline{\boldsymbol{\varphi}}_{i-1}t_{i-1})\left(h-\overline{h}\right)^T \right) \right\} \\ &- \left\{ \tilde{\pi}_i^{e} \left( (\boldsymbol{\varphi}-\overline{\boldsymbol{\varphi}})\left(ht-\overline{h}t+\overline{h}_{i-1}\Delta t_i\right)^T \right) + \tilde{\pi}_i^{e} \left( (\overline{\boldsymbol{\varphi}}t-\overline{\boldsymbol{\varphi}}_{i-1}t_{i-1})\left(h-\overline{h}\right)^T \right) \right\} \end{aligned} \right|^2 \right] \right)^{\frac{1}{2}}$$

$$\leq \left( E_P \left[ \left| \tilde{\pi}_i^{\prime e} \left( (\boldsymbol{\varphi}-\overline{\boldsymbol{\varphi}})\left(ht-\overline{h}t+\overline{h}_{i-1}\Delta t_i\right)^T \right) - \tilde{\pi}_i^{e} \left( (\boldsymbol{\varphi}-\overline{\boldsymbol{\varphi}})\left(ht-\overline{h}t+\overline{h}_{i-1}\Delta t_i\right)^T \right) \right|^2 \right] \right)^{\frac{1}{2}}$$

$$+ \left( E_P \left[ \left| \tilde{\pi}_i^{\prime e} \left( (\overline{\boldsymbol{\varphi}}t-\overline{\boldsymbol{\varphi}}_{i-1}t_{i-1})\left(h-\overline{h}\right)^T \right) - \tilde{\pi}_i^{e} \left( (\overline{\boldsymbol{\varphi}}t-\overline{\boldsymbol{\varphi}}_{i-1}t_{i-1})\left(h-\overline{h}\right)^T \right) \right|^2 \right] \right)^{\frac{1}{2}}$$

From Eqn. (A9) we write,

$$\left( E_P \left[ \left| \begin{aligned} &\left\{ \tilde{\pi}_i^{\prime e} \left( (\boldsymbol{\varphi}-\overline{\boldsymbol{\varphi}})\left(ht-\overline{h}t+\overline{h}_{i-1}\Delta t_i\right)^T \right) + \tilde{\pi}_i^{\prime e} \left( (\overline{\boldsymbol{\varphi}}t-\overline{\boldsymbol{\varphi}}_{i-1}t_{i-1})\left(h-\overline{h}\right)^T \right) \right\} \\ &- \left\{ \tilde{\pi}_i^{e} \left( (\boldsymbol{\varphi}-\overline{\boldsymbol{\varphi}})\left(ht-\overline{h}t+\overline{h}_{i-1}\Delta t_i\right)^T \right) + \tilde{\pi}_i^{e} \left( (\overline{\boldsymbol{\varphi}}t-\overline{\boldsymbol{\varphi}}_{i-1}t_{i-1})\left(h-\overline{h}\right)^T \right) \right\} \end{aligned} \right|^2 \right] \right)^{\frac{1}{2}}$$

$$\leq D_7 \left( \Delta t_i \frac{\left\| \mathrm{L}\left( (\boldsymbol{\varphi}-\overline{\boldsymbol{\varphi}})\left(ht-\overline{h}t+\overline{h}_{i-1}\Delta t_i\right)^T \right) \right\|}{\sqrt{N}} + \frac{\left\| (\boldsymbol{\varphi}-\overline{\boldsymbol{\varphi}})\left(ht-\overline{h}t+\overline{h}_{i-1}\Delta t_i\right)^T \right\|}{\sqrt{N}} \right.$$

$$\left. + \Delta t_i \frac{\left\| \mathrm{L}\left( (\overline{\boldsymbol{\varphi}}t-\overline{\boldsymbol{\varphi}}_{i-1}t_{i-1})\left(h-\overline{h}\right)^T \right) \right\|}{\sqrt{N}} + \frac{\left\| (\overline{\boldsymbol{\varphi}}t-\overline{\boldsymbol{\varphi}}_{i-1}t_{i-1})\left(h-\overline{h}\right)^T \right\|}{\sqrt{N}} \right)$$

(A12)

where, $D_7 > 0$ is a constant. The last term of inequality A11 may be written as:

$$\left( E_P \left[ \left| \begin{array}{l} \left\{ \alpha \tilde{\pi}_i'^e \left( (h-\bar{h})(h-\bar{h})^T \right) + (1-\alpha)\sigma_i^T \sigma_i \right\}^{-1} \\ -\left\{ \alpha \tilde{\pi}_i^e \left( (h-\bar{h})(h-\bar{h})^T \right) + (1-\alpha)\sigma_i^T \sigma_i \right\}^{-1} \end{array} \right|^2 \right] \right)^{\frac{1}{2}}$$

$$\leq \alpha \left( E_P \left[ \left| \begin{array}{l} \tilde{\pi}_i^e \left( (h-\bar{h})(h-\bar{h})^T \right) \\ -\tilde{\pi}_i'^e \left( (h-\bar{h})(h-\bar{h})^T \right) \end{array} \right|^2 \right] \right)^{\frac{1}{2}} \left( E_P \left[ \left| \begin{array}{l} \left\{ \alpha \tilde{\pi}_i'^e \left( (h-\bar{h})(h-\bar{h})^T \right) + (1-\alpha)\sigma_i^T \sigma_i \right\}^{-1} \\ \left\{ \alpha \tilde{\pi}_i^e \left( (h-\bar{h})(h-\bar{h})^T \right) + (1-\alpha)\sigma_i^T \sigma_i \right\}^{-1} \end{array} \right|^2 \right] \right)^{\frac{1}{2}}$$

$$\leq D_8 \alpha \left( \Delta t_i \frac{\left\| L\left( (h-\bar{h})(h-\bar{h})^T \right) \right\|}{\sqrt{N}} + \frac{\left\| (h-\bar{h})(h-\bar{h})^T \right\|}{\sqrt{N}} \right) \left\| \left\{ \alpha (h-\bar{h})^T (h-\bar{h}) + (1-\alpha)\sigma_i^T \sigma_i \right\}^{-1} \right\|^2$$

(A13)

$D_8 > 0$ is a constant. Using A11, A12 and A13, we may arrive at the following inequality:

$$\left(E_P\left[\left|\tilde{\mathbf{G}}_i'(\boldsymbol{\varphi})-\tilde{\mathbf{G}}_i(\boldsymbol{\varphi})\right|^2\right]\right)^{\frac{1}{2}}$$

$$\leq D_9 \begin{pmatrix} \Delta t_i \dfrac{\left\|\mathrm{L}\left((\boldsymbol{\varphi}-\bar{\boldsymbol{\varphi}})\left(ht-\bar{h}t+\bar{h}_{i-1}\Delta t_i\right)^T\right)\right\|}{\sqrt{N}} \\ +\dfrac{\left\|(\boldsymbol{\varphi}-\bar{\boldsymbol{\varphi}})\left(ht-\bar{h}t+\bar{h}_{i-1}\Delta t_i\right)^T\right\|}{\sqrt{N}} \\ +\Delta t_i \dfrac{\left\|\mathrm{L}\left((\bar{\boldsymbol{\varphi}}t-\bar{\boldsymbol{\varphi}}_{i-1}t_{i-1})(h-\bar{h})^T\right)\right\|}{\sqrt{N}} \\ +\dfrac{\left\|(\bar{\boldsymbol{\varphi}}t-\bar{\boldsymbol{\varphi}}_{i-1}t_{i-1})(h-\bar{h})^T\right\|}{\sqrt{N}} \end{pmatrix} E_P\left[\left|\left\{\alpha\tilde{\pi}_i'^e\left((h-\bar{h})(h-\bar{h})^T\right)\right\}^{-1}\right|^2\right]^{\frac{1}{2}}$$

$$+D_9\left(E_P\left[\left|\begin{cases}\tilde{\pi}_i^e\left((\boldsymbol{\varphi}-\bar{\boldsymbol{\varphi}})\left(ht-\bar{h}t+\bar{h}_{i-1}\Delta t_i\right)^T\right)\\+\tilde{\pi}_i^e\left((\bar{\boldsymbol{\varphi}}t-\bar{\boldsymbol{\varphi}}_{i-1}t_{i-1})(h-\bar{h})^T\right)\end{cases}\right|^2\right]\right)^{\frac{1}{2}}\alpha\begin{pmatrix}\Delta t_i \dfrac{\left\|\mathrm{L}\left((h-\bar{h})(h-\bar{h})^T\right)\right\|}{\sqrt{N}}\\+\dfrac{\left\|(h-\bar{h})(h-\bar{h})^T\right\|}{\sqrt{N}}\end{pmatrix}\left\|\left\{\alpha(h-\bar{h})^T(h-\bar{h})\right\}^{-1}\right\|^2$$

$D_9 > 0$ is a constant. This implies:

$$\left(E_P\left[\left|\tilde{\mathbf{G}}'_i(\boldsymbol{\varphi}) - \tilde{\mathbf{G}}_i(\boldsymbol{\varphi})\right|^2\right]\right)^{\frac{1}{2}}$$

$$\leq \frac{D_9}{\sqrt{N}} \begin{pmatrix} \Delta t_i \left\|\mathrm{L}\left((\boldsymbol{\varphi} - \bar{\boldsymbol{\varphi}})(ht - \bar{h}t + \bar{h}_{i-1}\Delta t_i)^T\right)\right\| \\ + \left\|(\boldsymbol{\varphi} - \bar{\boldsymbol{\varphi}})(ht - \bar{h}t + \bar{h}_{i-1}\Delta t_i)^T\right\| \\ + \Delta t_i \left\|\mathrm{L}\left((\bar{\boldsymbol{\varphi}}t - \bar{\boldsymbol{\varphi}}_{i-1}t_{i-1})(h - \bar{h})^T\right)\right\| \\ + \left\|(\bar{\boldsymbol{\varphi}}t - \bar{\boldsymbol{\varphi}}_{i-1}t_{i-1})(h - \bar{h})^T\right\| \end{pmatrix} \left\|\left\{\begin{array}{c}\alpha(h - \bar{h})^T(h - \bar{h}) \\ +(1-\alpha)\boldsymbol{\sigma}_i^T\boldsymbol{\sigma}_i\end{array}\right\}^{-1}\right\|$$

$$+ \frac{D_9}{\sqrt{N}} \left\{\begin{array}{c}\|(\boldsymbol{\varphi} - \bar{\boldsymbol{\varphi}})\|\|ht - \bar{h}t + \bar{h}_{i-1}\Delta t_i\| \\ + \|(\bar{\boldsymbol{\varphi}}t - \bar{\boldsymbol{\varphi}}_{i-1}t_{i-1})\|\|h - \bar{h}\|\end{array}\right\} \alpha \begin{pmatrix} \Delta t_i \left\|\mathrm{L}\left((h - \bar{h})(h - \bar{h})^T\right)\right\| \\ + \left\|(h - \bar{h})(h - \bar{h})^T\right\| \end{pmatrix} \left\|\left\{\begin{array}{c}\alpha(h - \bar{h})^T(h - \bar{h}) \\ +(1-\alpha)\boldsymbol{\sigma}_i^T\boldsymbol{\sigma}_i\end{array}\right\}^{-1}\right\|^2$$

(A14)

From Eqns. (A10) and (A14), we get:

$$\left(E_P\left[\left|\tilde{\mathbf{G}}'_i(\boldsymbol{\varphi})\tilde{I}'_i(\boldsymbol{\varphi})-\tilde{\mathbf{G}}_i(\boldsymbol{\varphi})\tilde{I}_i(\boldsymbol{\varphi})\right|^2\right]\right)^{\frac{1}{2}}$$

$$\leq \frac{D_{10}}{\sqrt{N}}\left(|Y_i|+\|h_i\|\right)\left\{ \begin{pmatrix} \Delta t_i \left\|\mathrm{L}\!\left((\boldsymbol{\varphi}-\overline{\boldsymbol{\varphi}})(ht-\overline{h}t+\overline{h}_{i-1}\Delta t_i)^T\right)\right\| \\ +\left\|(\boldsymbol{\varphi}-\overline{\boldsymbol{\varphi}})(ht-\overline{h}t+\overline{h}_{i-1}\Delta t_i)^T\right\| \\ +\Delta t_i \left\|\mathrm{L}\!\left((\overline{\boldsymbol{\varphi}}t-\overline{\boldsymbol{\varphi}}_{i-1}t_{i-1})(h-\overline{h})^T\right)\right\| \\ +\left\|(\overline{\boldsymbol{\varphi}}t-\overline{\boldsymbol{\varphi}}_{i-1}t_{i-1})(h-\overline{h})^T\right\| \end{pmatrix} \left\|\left\{\begin{matrix}\alpha(h-\overline{h})^T(h-\overline{h})\\ +(1-\alpha)\boldsymbol{\sigma}_i^T\boldsymbol{\sigma}_i\end{matrix}\right\}^{-1}\right\| \right.$$
$$+\left\{\begin{matrix}\|(\boldsymbol{\varphi}-\overline{\boldsymbol{\varphi}})\|\|ht-\overline{h}t+\overline{h}_{i-1}\Delta t_i\| \\ +\|(\overline{\boldsymbol{\varphi}}t-\overline{\boldsymbol{\varphi}}_{i-1}t_{i-1})\|\|h-\overline{h}\|\end{matrix}\right\}$$
$$\left. +\alpha\begin{pmatrix}\Delta t_i\left\|\mathrm{L}\!\left((h-\overline{h})(h-\overline{h})^T\right)\right\|\\ +\left\|(h-\overline{h})(h-\overline{h})^T\right\|\end{pmatrix}\left\|\left\{\begin{matrix}\alpha(h-\overline{h})^T(h-\overline{h})\\ +(1-\alpha)\boldsymbol{\sigma}_i^T\boldsymbol{\sigma}_i\end{matrix}\right\}^{-1}\right\|^2 \right\}$$

$$+\frac{D_{10}}{\sqrt{N}}\left\|\tilde{\mathbf{G}}_i(\boldsymbol{\varphi})\right\|\left(\Delta t_i\|\mathrm{L}(h)\|+\|h\|\right)$$

(A15)

Finally, from Eqn. A8, we get an error bound due to the ensemble approximation of the filtered estimate at time $t_i$:

$$\left(E_P\left[\left|\pi_i^{\prime e}(\boldsymbol{\varphi})-\pi_i^e(\boldsymbol{\varphi})\right|^2\right]\right)^{\frac{1}{2}}$$

$$\leq \frac{D_{11}}{\sqrt{N}}\left(|Y_i|+\|h_i\|\right)\left\{\begin{array}{l}\left(\begin{array}{l}\Delta t_i\left\|\mathrm{L}\left((\boldsymbol{\varphi}-\overline{\boldsymbol{\varphi}})\left(ht-\overline{h}t+\overline{h}_{i-1}\Delta t_i\right)^T\right)\right\| \\ +\left\|(\boldsymbol{\varphi}-\overline{\boldsymbol{\varphi}})\left(ht-\overline{h}t+\overline{h}_{i-1}\Delta t_i\right)^T\right\| \\ +\Delta t_i\left\|\mathrm{L}\left((\overline{\boldsymbol{\varphi}}t-\overline{\boldsymbol{\varphi}}_{i-1}t_{i-1})(h-\overline{h})^T\right)\right\| \\ +\left\|(\overline{\boldsymbol{\varphi}}t-\overline{\boldsymbol{\varphi}}_{i-1}t_{i-1})(h-\overline{h})^T\right\|\end{array}\right)\left\|\left\{\begin{array}{l}\alpha(h-\overline{h})^T(h-\overline{h}) \\ +(1-\alpha)\boldsymbol{\sigma}_i^T\boldsymbol{\sigma}_i\end{array}\right\}^{-1}\right\| \\ +\left\{\begin{array}{l}\|(\boldsymbol{\varphi}-\overline{\boldsymbol{\varphi}})\|\|ht-\overline{h}t+\overline{h}_{i-1}\Delta t_i\| \\ +\|(\overline{\boldsymbol{\varphi}}t-\overline{\boldsymbol{\varphi}}_{i-1}t_{i-1})\|\|h-\overline{h}\|\end{array}\right\} \\ \alpha\left(\begin{array}{l}\Delta t_i\left\|\mathrm{L}\left((h-\overline{h})(h-\overline{h})^T\right)\right\| \\ +\left\|(h-\overline{h})(h-\overline{h})^T\right\|\end{array}\right)\left\|\left\{\begin{array}{l}\alpha(h-\overline{h})^T(h-\overline{h}) \\ +(1-\alpha)\boldsymbol{\sigma}_i^T\boldsymbol{\sigma}_i\end{array}\right\}^{-1}\right\|^2\end{array}\right\}$$

$$+\frac{D_{11}}{\sqrt{N}}\left(\left\|\tilde{\mathbf{G}}_i(\boldsymbol{\varphi})\right\|\left(\Delta t_i\|\mathrm{L}(h)\|+\|h\|\right)+\left(\Delta t_i\|\mathrm{L}(\boldsymbol{\varphi})\|+\|\boldsymbol{\varphi}\|\right)\right)$$

(A16)

Here, $D_{12} > 0$ is a constant. Upon simplification of Eqn. (A16), we get:

$$\left(E_P\left[\left|\pi_i^{\prime e}(\boldsymbol{\varphi})-\pi_i^e(\boldsymbol{\varphi})\right|^2\right]\right)^{\frac{1}{2}}$$

$$\leq \frac{D_{12}}{\sqrt{N}}(|Y_i|+\|h_i\|)\left\{\begin{pmatrix}\Delta t_i\left\|\mathrm{L}\left((\boldsymbol{\varphi}-\overline{\boldsymbol{\varphi}})\left(ht-\overline{h}t+\overline{h}_{i-1}\Delta t_i\right)^T\right)\right\|\\+\|\boldsymbol{\varphi}\|\|h\|\Delta t_i\\+\Delta t_i\left\|\mathrm{L}\left((\overline{\boldsymbol{\varphi}}t-\overline{\boldsymbol{\varphi}}_{i-1}t_{i-1})(h-\overline{h})^T\right)\right\|\\+\|\boldsymbol{\varphi}\|\|h\|\Delta t_i\end{pmatrix}\left\|\left\{\begin{matrix}\alpha(h-\overline{h})^T(h-\overline{h})\\+(1-\alpha)\boldsymbol{\sigma}_i^T\boldsymbol{\sigma}_i\end{matrix}\right\}^{-1}\right\|\\+\|\boldsymbol{\varphi}\|\|h\|\Delta t_i\alpha\begin{pmatrix}\Delta t_i\left\|\mathrm{L}\left((h-\overline{h})(h-\overline{h})^T\right)\right\|\\+\left\|(h-\overline{h})(h-\overline{h})^T\right\|\end{pmatrix}\left\|\left\{\begin{matrix}\alpha(h-\overline{h})^T(h-\overline{h})\\+(1-\alpha)\boldsymbol{\sigma}_i^T\boldsymbol{\sigma}_i\end{matrix}\right\}^{-1}\right\|^2\end{pmatrix}\right\}$$

$$+\frac{D_{11}}{\sqrt{N}}\left(\left\|\tilde{\mathbf{G}}_i(\boldsymbol{\varphi})\right\|(\Delta t_i\|\mathrm{L}(h)\|+\|h\|)+(\Delta t_i\|\mathrm{L}(\boldsymbol{\varphi})\|+\|\boldsymbol{\varphi}\|)\right) \quad (A17)$$

where $D_{12} > 0$ is a constant. Simplification of the inequality in (A17) yields:

$$\left(E_P\left[\left|\pi_i^{\prime e}(\boldsymbol{\varphi})-\pi_i^e(\boldsymbol{\varphi})\right|^2\right]\right)^{\frac{1}{2}}$$

$$\leq \frac{D''}{\sqrt{N}}\Delta t_i(|Y_i|+\|h_i\|)\left\{\begin{pmatrix}\left\|\mathrm{L}\left((\boldsymbol{\varphi}-\overline{\boldsymbol{\varphi}})\left(ht-\overline{h}t+\overline{h}_{i-1}\Delta t_i\right)^T\right)\right\|\\+\|\boldsymbol{\varphi}\|\|h\|+\left\|\mathrm{L}\left((\overline{\boldsymbol{\varphi}}t-\overline{\boldsymbol{\varphi}}_{i-1}t_{i-1})(h-\overline{h})^T\right)\right\|\end{pmatrix}\left\|\left\{\begin{matrix}\alpha(h-\overline{h})^T(h-\overline{h})\\+(1-\alpha)\boldsymbol{\sigma}_i^T\boldsymbol{\sigma}_i\end{matrix}\right\}^{-1}\right\|\\+\|\boldsymbol{\varphi}\|\|h\|\alpha\left(\Delta t_i\left\|\mathrm{L}\left((h-\overline{h})(h-\overline{h})^T\right)\right\|+\left\|(h-\overline{h})(h-\overline{h})^T\right\|\right)\left\|\left\{\begin{matrix}\alpha(h-\overline{h})^T(h-\overline{h})\\+(1-\alpha)\boldsymbol{\sigma}_i^T\boldsymbol{\sigma}_i\end{matrix}\right\}^{-1}\right\|^2\end{pmatrix}\right\}$$

$$+\frac{D''}{\sqrt{N}}\left(\left\|\tilde{\mathbf{G}}_i(\boldsymbol{\varphi})\right\|(\Delta t_i\|\mathrm{L}(h)\|+\|h\|)+(\Delta t_i\|\mathrm{L}(\boldsymbol{\varphi})\|+\|\boldsymbol{\varphi}\|)\right)$$

(A18)

$D'' > 0$ is a constant.

□

*Proof of Theorem 1:*

Combining Lemmas 1, 2 and Eqn. (A1), we may arrive at the following error bound for the filtered estimate.

$$\left(E_P\left[\left|\pi_i(\varphi) - \pi_i'^e(\varphi)\right|^2\right]\right)^{\frac{1}{2}} \leq D'(\Delta t_i)^{\frac{1}{2}}$$

$$+ \frac{D''}{\sqrt{N}} \Delta t_i \left(|Y_i| + \|h_i\|\right) \left\{ \begin{bmatrix} \left\|L\left((\varphi - \overline{\varphi})(ht - \overline{h}t + \overline{h}_{i-1}\Delta t_i)^T\right)\right\| \\ + \|\varphi\|\|h\| + \left\|L\left((\overline{\varphi}t - \overline{\varphi}_{i-1}t_{i-1})(h-\overline{h})^T\right)\right\| \end{bmatrix} \left\|\begin{Bmatrix}\alpha(h-\overline{h})^T(h-\overline{h}) \\ + (1-\alpha)\sigma_i^T\sigma_i\end{Bmatrix}^{-1}\right\| \\ + \|\varphi\|\|h\|\alpha\left(\Delta t_i \left\|L\left((h-\overline{h})(h-\overline{h})^T\right)\right\| + \left\|(h-\overline{h})(h-\overline{h})^T\right\|\right) \left\|\begin{Bmatrix}\alpha(h-\overline{h})^T(h-\overline{h}) \\ + (1-\alpha)\sigma_i^T\sigma_i\end{Bmatrix}^{-1}\right\|^2 \right\}$$

$$+ \frac{D''}{\sqrt{N}} \left(\|\tilde{G}_i(\varphi)\|(\Delta t_i\|L(h)\| + \|h\|) + (\Delta t_i\|L(\varphi)\| + \|\varphi\|)\right)$$

(A19)

□

*Proof of Theorem 2:*

The prediction equation is taken as:

$$\tilde{\varphi}_t = \varphi_{i-1} + \int_{t_{i-1}}^t L(\varphi_s)ds + \int_{t_{i-1}}^t \varphi_s' f_s dB_s \tag{A20}$$

$$\Rightarrow \|\tilde{\varphi}_i\|_{2,(v_{t_i} \times P)} \leq \|\varphi_{i-1}\|_{2,(v_{t_i} \times P)} + \left\|\int_{t_{i-1}}^{t_i} L(\varphi_s)ds\right\|_{2,(v_{t_i} \times P)} + \left\|\int_{t_{i-1}}^{t_i} \varphi_s' f_s dB_s\right\|_{2,(v_{t_i} \times P)} \tag{A21}$$

where, $v_t$ is a Lebesgue measure in $t$ and $\|\cdot\|_{2,(v_{t_i} \times P)}$ is a standard matrix norm, e.g. the 2-norm [39]. The first term on the RHS of Eqn. (A21) satisfies:

$$\|\varphi_{i-1}\|_{2,(v_{t_i} \times P)} = \sqrt{\Delta t}\|\varphi_{i-1}\|_{2,P} < \infty \tag{A22}$$

For the second term on the RHS of (A21), using Jensen's inequality ( $\left(t\int_0^t a_s ds\right)^2 \leq t^{-1}\int_0^t a_s^2 ds$ ) and linear growth property of $L(\varphi_t)$, we may write:

$$\left\|\int_{t_{i-1}}^{t_i} L(\varphi_s)ds\right\|^2_{2,(v_{t_i}\times P)} \leq (\Delta t_i)^2 \|L(\varphi)\|^2_{2,(v_{t_i}\times P)} \leq (\Delta t)^2 \left(1+\|\varphi\|_{2,(v_{t_i}\times P)}\right)^2 < \infty \tag{A22}$$

Similarly, for the third term on the RHS of (A21), using Ito's isometry and linear growth property of $\varphi'_t f_t$, we may write:

$$\left\|\int_{t_{i-1}}^{t_i} \varphi'_s f_s dB_s\right\|^2_{2,(v_{t_i}\times P)} \leq (\Delta t_i)\|\varphi' f\|^2_{2,(v_{t_i}\times P)} \leq (\Delta t_i)\left(1+\|\varphi\|_{2,(v_{t_i}\times P)}\right)^2 < \infty$$

$$\tag{A23}$$

Hence we conclude that $\tilde{\varphi}_i \in L^2(v_{t_i}\times P)$. This paves the way for proving the existence of a limiting solution for the inner iteration at a given time $t_i$. The iterated update equation corresponding to the conditioned process $\varphi_t$ may be written as:

$$\varphi_{t,k} = \varphi_{t,k-1} + \beta_{k-1}G_{t,k-1}\{Y_t - h_{t,k-1}\}, \quad k=1,2,... \tag{A24}$$

For further work, the update function is denoted as:

$$\aleph_{t,k-1}(\varphi) := \varphi_{t,k-1} + \beta_{k-1}G_{t,k-1}\{Y_t - h_{t,k-1}\} \tag{A25}$$

From (A25) we may write:

$$\|\aleph_{t,k-1}(\varphi)\|_{2,(v_{t_i}\times P)} \leq \|\varphi_{t,k-1}\|_{2,(v_{t_i}\times P)} + \|\beta_{k-1}G_{t,k-1}\{Y_t - h_{t,k-1}\}\|_{2,(v_{t_i}\times P)} \tag{A26}$$

Defining

$$\|\aleph_t(\varphi)\|_{2,(v_{t_i}\times P)} \leq \|\varphi_t\|_{2,(v_{t_i}\times P)} + \|G_t\{Y_t - h_t\}\|_{2,(v_{t_i}\times P)} \tag{A27}$$

we have:

$$\|\aleph_t(\varphi)\|_{2,(v_{t_i} \times P)} \leq \|\varphi_t\|_{2,(v_{t_i} \times P)} + \|G_t\|_{2,(v_{t_i} \times P)} \|\{Y_t - h_t\}\|_{2,(v_{t_i} \times P)} \tag{A28}$$

Now, recalling that $\varphi_t \in C_b^2$ and $h$ satisfy linear growth property, we may conclude from (A28) that $\|\aleph_t(\varphi)\|_{2,(v_{t_i} \times P)} \leq \infty$ which implies that $\aleph_t$ indeed maps to an $\mathcal{F}_t^Y$ adapted process in $L^2(v_{t_i} \times P)$.

Next, we show that $\aleph_t$ is a continuous map, i.e., as $\|(\varphi_{t,k-1} - \varphi_t)\|_{2,(v_{t_i} \times P)} \to 0$ and $\beta_{k-1} \to 1$,

$$\|\aleph_{t,k-1}(\varphi) - \aleph_t(\varphi)\|_{2,(v_{t_i} \times P)} \to 0$$

Taking norm of $\aleph_{t,k-1}(\varphi) - \aleph_t(\varphi)$ we get:

$$\|\aleph_{t,k-1}(\varphi) - \aleph_t(\varphi)\|_{2,(v_{t_i} \times P)} \leq \|\varphi_{t,k-1} - \varphi\|_{2,(v_{t_i} \times P)} + \|\beta_{k-1} G_{t,k-1}\{Y_t - h_{t,k-1}\} - G_t\{Y_t - h_t\}\|_{2,(v_{t_i} \times P)}$$

$$\tag{A29}$$

For expositional ease, we replace $\|\cdot\|_{2,(v_{t_i} \times P)}$ by $\|\cdot\|$. The second term on the RHS of (A29) may be split as:

$$\|\beta_{k-1} G_{t,k-1}\{Y_t - h_{t,k-1}\} - G_t\{Y_t - h_t\}\| \leq |\beta_{k-1} - 1| \|G_{t,k-1}\{Y_t - h_{t,k-1}\}\|$$
$$+ \|G_{t,k-1}\{Y_t - h_{t,k-1}\} - G_t\{Y_t - h_t\}\|$$
$$\leq |\beta_{k-1} - 1| \|G_{t,k-1}\{Y_t - h_{t,k-1}\}\|$$
$$+ \|G_{t,k-1} - G_t\| \|\{Y_t - h_{t,k-1}\}\|$$
$$+ \|G_t\| \|h_t - h_{t,k-1}\|$$

$$\tag{A30}$$

The factor $\|\mathbf{G}_{t,k-1} - \mathbf{G}_t\|$ in the second term on the RHS of (A30) may be split as:

$$\|\mathbf{G}_{t,k-1} - \mathbf{G}_t\| \leq \left\| \begin{array}{l} \frac{1}{N} \left\{ \begin{array}{l} \left(\mathbf{\Phi}_{t,k-1} - \hat{\mathbf{\Phi}}_{t,k-1}\right)\left(\mathbf{H}_{t,k-1}{}^T t - \hat{\mathbf{H}}_{i-1}{}^T t_{i-1} - \Delta\hat{\mathbf{H}}_{t,k-1}{}^T t\right) \\ + \left(\hat{\mathbf{\Phi}}_{t,k-1} t - \hat{\mathbf{\Phi}}_{i-1} t_{i-1}\right)\left(\mathbf{H}_{t,k-1}{}^T - \hat{\mathbf{H}}_{t,k-1}{}^T\right) \end{array} \right\} \\ \left\{ \alpha \frac{1}{N-1}\left(\mathbf{H}_{t,k-1} - \hat{\mathbf{H}}_{t,k-1}\right)\left(\mathbf{H}_{t,k-1}{}^T - \hat{\mathbf{H}}_{t,k-1}^T\right) + (1-\alpha)\boldsymbol{\sigma}_t{}^T\boldsymbol{\sigma}_t \right\}^{-1} \\ -\frac{1}{N}\left\{\left(\mathbf{\Phi}_t - \hat{\mathbf{\Phi}}_t\right)\left(\mathbf{H}_t^T t - \hat{\mathbf{H}}_{i-1}^T t_{i-1} - \Delta\hat{\mathbf{H}}_t^T t\right) + \left(\hat{\mathbf{\Phi}}_t t - \hat{\mathbf{\Phi}}_{i-1} t_{i-1}\right)\left(\mathbf{H}_t^T - \hat{\mathbf{H}}_t^T\right)\right\} \\ \left\{ \alpha \frac{1}{N-1}\left(\mathbf{H}_t - \hat{\mathbf{H}}_t\right)\left(\mathbf{H}_t^T - \hat{\mathbf{H}}_t^T\right) + (1-\alpha)\boldsymbol{\sigma}_t{}^T\boldsymbol{\sigma}_t \right\}^{-1} \end{array} \right\|$$

$$\leq \left\| \begin{array}{l} \frac{1}{N} \left\{ \begin{array}{l} \left(\mathbf{\Phi}_{t,k-1} - \hat{\mathbf{\Phi}}_{t,k-1}\right)\left(\mathbf{H}_{t,k-1}{}^T t - \hat{\mathbf{H}}_{i-1}{}^T t_{i-1} - \Delta\hat{\mathbf{H}}_{t,k-1}{}^T t\right) \\ + \left(\hat{\mathbf{\Phi}}_{t,k-1} t - \hat{\mathbf{\Phi}}_{i-1} t_{i-1}\right)\left(\mathbf{H}_{t,k-1}{}^T - \hat{\mathbf{H}}_{t,k-1}{}^T\right) \end{array} \right\} \\ -\frac{1}{N}\left\{\left(\mathbf{\Phi}_t - \hat{\mathbf{\Phi}}_t\right)\left(\mathbf{H}_t^T t - \hat{\mathbf{H}}_{i-1}^T t_{i-1} - \Delta\hat{\mathbf{H}}_t^T t\right) + \left(\hat{\mathbf{\Phi}}_t t - \hat{\mathbf{\Phi}}_{i-1} t_{i-1}\right)\left(\mathbf{H}_t^T - \hat{\mathbf{H}}_t^T\right)\right\} \end{array} \right\|$$

$$\left\| \left\{ \alpha \frac{1}{N-1}\left(\mathbf{H}_{t,k-1} - \hat{\mathbf{H}}_{t,k-1}\right)\left(\mathbf{H}_{t,k-1}{}^T - \hat{\mathbf{H}}_{t,k-1}^T\right) + (1-\alpha)\boldsymbol{\sigma}_t{}^T\boldsymbol{\sigma}_t \right\}^{-1} \right\|$$

$$+ \left\| \frac{1}{N}\left\{\left(\mathbf{\Phi}_t - \hat{\mathbf{\Phi}}_t\right)\left(\mathbf{H}_t^T t - \hat{\mathbf{H}}_{i-1}^T t_{i-1} - \Delta\hat{\mathbf{H}}_t^T t\right) + \left(\hat{\mathbf{\Phi}}_t t - \hat{\mathbf{\Phi}}_{i-1} t_{i-1}\right)\left(\mathbf{H}_t^T - \hat{\mathbf{H}}_t^T\right)\right\} \right\|$$

$$\left\| \begin{array}{l} \left\{ \alpha \frac{1}{N-1}\left(\mathbf{H}_{t,k-1} - \hat{\mathbf{H}}_{t,k-1}\right)\left(\mathbf{H}_{t,k-1}{}^T - \hat{\mathbf{H}}_{t,k-1}^T\right) + (1-\alpha)\boldsymbol{\sigma}_t{}^T\boldsymbol{\sigma}_t \right\}^{-1} \\ -\left\{ \alpha \frac{1}{N-1}\left(\mathbf{H}_t - \hat{\mathbf{H}}_t\right)\left(\mathbf{H}_t^T - \hat{\mathbf{H}}_t^T\right) + (1-\alpha)\boldsymbol{\sigma}_t{}^T\boldsymbol{\sigma}_t \right\}^{-1} \end{array} \right\|$$

(A31)

From the first term on the RHS of (A31) we take the following term and split it as:

$$\left\|\frac{1}{N}\left\{\begin{array}{l}\left(\mathbf{\Phi}_{t,k-1}-\widehat{\mathbf{\Phi}}_{t,k-1}\right)\left(\mathbf{H}_{t,k-1}{}^T t-\widehat{\mathbf{H}}_{i-1}{}^T t_{i-1}-\Delta\widehat{\mathbf{H}}_{t,k-1}{}^T t\right)\\+\left(\widehat{\mathbf{\Phi}}_{t,k-1}t-\widehat{\mathbf{\Phi}}_{i-1}t_{i-1}\right)\left(\mathbf{H}_{t,k-1}{}^T-\widehat{\mathbf{H}}_{t,k-1}{}^T\right)\end{array}\right\}\right.$$
$$\left.-\frac{1}{N}\left\{\left(\mathbf{\Phi}_t-\widehat{\mathbf{\Phi}}_t\right)\left(\mathbf{H}_t^T t-\widehat{\mathbf{H}}_{i-1}{}^T t_{i-1}-\Delta\widehat{\mathbf{H}}_t^T t\right)+\left(\widehat{\mathbf{\Phi}}_t t-\widehat{\mathbf{\Phi}}_{i-1}t_{i-1}\right)\left(\mathbf{H}_t^T-\widehat{\mathbf{H}}_t^T\right)\right\}\right\|$$
$$\leq \frac{1}{N}\left\|\left(\mathbf{\Phi}_{t,k-1}-\widehat{\mathbf{\Phi}}_{t,k-1}\right)-\left(\mathbf{\Phi}_t-\widehat{\mathbf{\Phi}}_t\right)\right\|\left\|\left(\mathbf{H}_{t,k-1}{}^T t-\widehat{\mathbf{H}}_{i-1}{}^T t_{i-1}-\Delta\widehat{\mathbf{H}}_{t,k-1}{}^T t\right)\right\|$$
$$+\frac{1}{N}\left\|\left(\mathbf{\Phi}_t-\widehat{\mathbf{\Phi}}_t\right)\right\|\left\|\begin{array}{c}\left(\mathbf{H}_{t,k-1}{}^T t-\widehat{\mathbf{H}}_{i-1}{}^T t_{i-1}-\Delta\widehat{\mathbf{H}}_{t,k-1}{}^T t\right)\\-\left(\mathbf{H}_t^T t-\widehat{\mathbf{H}}_{i-1}{}^T t_{i-1}-\Delta\widehat{\mathbf{H}}_t^T t\right)\end{array}\right\|$$
$$+\frac{1}{N}\left(\begin{array}{c}\left\|\left(\widehat{\mathbf{\Phi}}_{t,k-1}t-\widehat{\mathbf{\Phi}}_{i-1}t_{i-1}\right)-\left(\widehat{\mathbf{\Phi}}_t t-\widehat{\mathbf{\Phi}}_{i-1}t_{i-1}\right)\right\|\left\|\left(\mathbf{H}_{t,k-1}{}^T-\widehat{\mathbf{H}}_{t,k-1}{}^T\right)\right\|\\+\left\|\left(\widehat{\mathbf{\Phi}}_t t-\widehat{\mathbf{\Phi}}_{i-1}t_{i-1}\right)\right\|\left\|\left(\mathbf{H}_{t,k-1}{}^T-\widehat{\mathbf{H}}_{t,k-1}{}^T\right)-\left(\mathbf{H}_t^T-\widehat{\mathbf{H}}_t^T\right)\right\|\end{array}\right)$$
$$\leq \frac{D_{13}}{N}\left\|\left(\boldsymbol{\varphi}_{t,k-1}-\boldsymbol{\varphi}_t\right)\right\|\left\|\left(\mathbf{H}_{t,k-1}{}^T t-\widehat{\mathbf{H}}_{i-1}{}^T t_{i-1}-\Delta\widehat{\mathbf{H}}_{t,k-1}{}^T t\right)\right\|$$
$$+\frac{D_{13}}{N}t\left\|\left(\mathbf{\Phi}_t-\widehat{\mathbf{\Phi}}_t\right)\right\|\left\|\left(h_t-h_{t,k-1}\right)\right\|$$
$$+\frac{1}{N}\left(t\left\|\left(\boldsymbol{\varphi}_{t,k-1}-\boldsymbol{\varphi}_t\right)\right\|\left\|\left(\mathbf{H}_{t,k-1}{}^T-\widehat{\mathbf{H}}_{t,k-1}{}^T\right)\right\|+\left\|\left(\widehat{\mathbf{\Phi}}_t t-\widehat{\mathbf{\Phi}}_{i-1}t_{i-1}\right)\right\|\left\|\left(h_t-h_{t,k-1}\right)\right\|\right)$$

(A32)

where the constant $D_{13} > 0$. From the last term on the RHS of (A31) we take the following term and split it as:

$$\left\|\left\{\alpha\frac{1}{N-1}\left(\mathbf{H}_{t,k-1}-\widehat{\mathbf{H}}_{t,k-1}\right)\left(\mathbf{H}_{t,k-1}{}^T-\widehat{\mathbf{H}}_{t,k-1}^T\right)+(1-\alpha)\boldsymbol{\sigma}_t^T\boldsymbol{\sigma}_t\right\}^{-1}\right.$$
$$\left.-\left\{\alpha\frac{1}{N-1}\left(\mathbf{H}_t-\widehat{\mathbf{H}}_t\right)\left(\mathbf{H}_t^T-\widehat{\mathbf{H}}_t^T\right)+(1-\alpha)\boldsymbol{\sigma}_t^T\boldsymbol{\sigma}_t\right\}^{-1}\right\|$$
$$\leq \alpha\frac{1}{N-1}\left(\left\|\left\{\left(\mathbf{H}_t-\widehat{\mathbf{H}}_t\right)\left(\mathbf{H}_t^T-\widehat{\mathbf{H}}_t^T\right)-\left(\mathbf{H}_{t,k-1}-\widehat{\mathbf{H}}_{t,k-1}\right)\left(\mathbf{H}_{t,k-1}{}^T-\widehat{\mathbf{H}}_{t,k-1}^T\right)\right\}\right\|\right) \quad (A33)$$
$$\left\|\left\{\alpha\frac{1}{N-1}\left(\mathbf{H}_{t,k-1}-\widehat{\mathbf{H}}_{t,k-1}\right)\left(\mathbf{H}_{t,k-1}{}^T-\widehat{\mathbf{H}}_{t,k-1}^T\right)+(1-\alpha)\boldsymbol{\sigma}_t^T\boldsymbol{\sigma}_t\right\}^{-1}\right\|$$
$$\left\|\left\{\alpha\frac{1}{N-1}\left(\mathbf{H}_t-\widehat{\mathbf{H}}_t\right)\left(\mathbf{H}_t^T-\widehat{\mathbf{H}}_t^T\right)+(1-\alpha)\boldsymbol{\sigma}_t^T\boldsymbol{\sigma}_t\right\}^{-1}\right\|$$

Splitting the first term on the RHS of (A33):

$$\alpha \frac{1}{N-1} \left( \left\| \left\{ \left( \mathbf{H}_t - \hat{\mathbf{H}}_t \right) \left( \mathbf{H}_t^T - \hat{\mathbf{H}}_t^T \right) - \left( \mathbf{H}_{t,k-1} - \hat{\mathbf{H}}_{t,k-1} \right) \left( \mathbf{H}_{t,k-1}^T - \hat{\mathbf{H}}_{t,k-1}^T \right) \right\} \right\| \right)$$

$$\leq \alpha \frac{1}{N-1} \left( \left( \left\| \left( \mathbf{H}_t - \mathbf{H}_{t,k-1} \right) \right\| + \left\| \left( \hat{\mathbf{H}}_{t,k-1} - \hat{\mathbf{H}}_t \right) \right\| \right) \left\| \left( \mathbf{H}_t^T - \hat{\mathbf{H}}_t^T \right) \right\| \right)$$

$$+ \alpha \frac{1}{N-1} \left( \left\| \left( \mathbf{H}_{t,k-1} - \hat{\mathbf{H}}_{t,k-1} \right) \right\| \left( \left\| \left( \mathbf{H}_t^T - \mathbf{H}_{t,k-1}^T \right) \right\| + \left\| \left( \hat{\mathbf{H}}_{t,k-1}^T - \hat{\mathbf{H}}_t^T \right) \right\| \right) \right)$$

$$\leq D_{14} \alpha \frac{1}{N-1} \left( \left\| \left( h_t - h_{t,k-1} \right) \right\| \left\| \left( \mathbf{H}_t^T - \hat{\mathbf{H}}_t^T \right) \right\| + \left\| \left( \mathbf{H}_{t,k-1} - \hat{\mathbf{H}}_{t,k-1} \right) \right\| \left\| \left( h_t - h_{t,k-1} \right) \right\| \right)$$

(A34)

where the constant $D_{14} > 0$. From (A31-34), we have:

$$\left\| \mathbf{G}_{t,k-1} - \mathbf{G}_t \right\| \leq D_{15} \left\{ \begin{array}{l} \frac{1}{N} \left\| \left( \boldsymbol{\varphi}_{t,k-1} - \boldsymbol{\varphi}_t \right) \right\| \left\| \left( \mathbf{H}_{t,k-1}^T t - \hat{\mathbf{H}}_{i-1}^T t_{i-1} - \Delta \hat{\mathbf{H}}_{t,k-1}^T t \right) \right\| \\ + \frac{1}{N} t \left\| \left( \boldsymbol{\Phi}_t - \hat{\boldsymbol{\Phi}}_t \right) \right\| \left\| \left( h_t - h_{t,k-1} \right) \right\| \\ + \frac{1}{N} \left( t \left\| \left( \boldsymbol{\varphi}_{t,k-1} - \boldsymbol{\varphi}_t \right) \right\| \left\| \left( \mathbf{H}_{t,k-1}^T - \hat{\mathbf{H}}_{t,k-1}^T \right) \right\| + \left\| \left( \hat{\boldsymbol{\Phi}}_t t - \hat{\boldsymbol{\Phi}}_{i-1} t_{i-1} \right) \right\| \left\| \left( h_t - h_{t,k-1} \right) \right\| \right) \end{array} \right\}$$

$$\left\| \left\{ \alpha \frac{1}{N-1} \left( \mathbf{H}_{t,k-1} - \hat{\mathbf{H}}_{t,k-1} \right) \left( \mathbf{H}_{t,k-1}^T - \hat{\mathbf{H}}_{t,k-1}^T \right) + (1-\alpha) \boldsymbol{\sigma}_t^T \boldsymbol{\sigma}_t \right\}^{-1} \right\|$$

$$+ D_{15} \left\| \frac{1}{N} \left\{ \left( \boldsymbol{\Phi}_t - \hat{\boldsymbol{\Phi}}_t \right) \left( \mathbf{H}_t^T t - \hat{\mathbf{H}}_{i-1}^T t_{i-1} - \Delta \hat{\mathbf{H}}_t^T t \right) + \left( \hat{\boldsymbol{\Phi}}_t t - \hat{\boldsymbol{\Phi}}_{i-1} t_{i-1} \right) \left( \mathbf{H}_t^T - \hat{\mathbf{H}}_t^T \right) \right\} \right\|$$

$$\left\{ \begin{array}{l} \left( \alpha \frac{1}{N-1} \left( \left\| \left( h_t - h_{t,k-1} \right) \right\| \left\| \left( \mathbf{H}_t^T - \hat{\mathbf{H}}_t^T \right) \right\| + \left\| \left( \mathbf{H}_{t,k-1} - \hat{\mathbf{H}}_{t,k-1} \right) \right\| \left\| \left( h_t - h_{t,k-1} \right) \right\| \right) \right) \\ \left\| \left\{ \alpha \frac{1}{N-1} \left( \mathbf{H}_{t,k-1} - \hat{\mathbf{H}}_{t,k-1} \right) \left( \mathbf{H}_{t,k-1}^T - \hat{\mathbf{H}}_{t,k-1}^T \right) + (1-\alpha) \boldsymbol{\sigma}_t^T \boldsymbol{\sigma}_t \right\}^{-1} \right\| \\ \left\| \left\{ \alpha \frac{1}{N-1} \left( \mathbf{H}_t - \hat{\mathbf{H}}_t \right) \left( \mathbf{H}_t^T - \hat{\mathbf{H}}_t^T \right) + (1-\alpha) \boldsymbol{\sigma}_t^T \boldsymbol{\sigma}_t \right\}^{-1} \right\| \end{array} \right\}$$

(A35)

Since $h_t$ is Lipschitz continuous, from (A35) we may conclude:

$$\|\mathbf{G}_{t,k-1} - \mathbf{G}_t\| \to 0$$

as $\|(\boldsymbol{\varphi}_{t,k-1} - \boldsymbol{\varphi}_t)\| \to 0$ and $|\beta_{k-1} - 1| \to 0$

which implies via (A29) and (A30):

$$\|\aleph_{t,k-1}(\boldsymbol{\varphi}) - \aleph_t(\boldsymbol{\varphi})\| \to 0 \tag{A36}$$

as $\|(\boldsymbol{\varphi}_{t,k-1} - \boldsymbol{\varphi}_t)\| \to 0$ and $|\beta_{k-1} - 1| \to 0$.

Now consider the sequence $\{\boldsymbol{\varphi}_{t,k+1} = \aleph_t(\boldsymbol{\varphi}_k) = \aleph^{k-1}(\boldsymbol{\varphi}_{t,0})\}_{k=1}^{\infty}$ where the iterations are of the stochastic Picard's form. If, for a given $t$, $\{\boldsymbol{\varphi}_{t,k}\}_{k=1}^{K}$ is a Cauchy sequence in $L^2(\nu_{t_i} \times P)$, it must converge to some $\mathcal{F}_t^Y$ process $\boldsymbol{\varphi}_t$ in $L^2(\nu_{t_i} \times P)$, i.e. $\boldsymbol{\varphi}_t$ must be a fixed point in $L^2(\nu_{t_i} \times P)$.

By continuity of $\aleph_t$, $\aleph_t(\boldsymbol{\varphi}_k) \to \aleph_t(\boldsymbol{\varphi})$. But $\aleph_t(\boldsymbol{\varphi}_k) = \boldsymbol{\varphi}_{k+1} \to \boldsymbol{\varphi}_t$. Hence, $\aleph_t(\boldsymbol{\varphi}) = \boldsymbol{\varphi}_t$ which means that $\boldsymbol{\varphi}_t$ is the desired solution at time $t$ arrived at through inner iterations. In other words, the aim of driving $\{Y_t - h_t\}$ to a zero-mean martingale is met. Hence, the remaining job is to show that $\{\boldsymbol{\varphi}_{t,k}\}$ is indeed a Cauchy sequence in $L^2(\nu_{t_i} \times P)$.

In the same way as the inequality (A28) is arrived at, we may have:

$$\|\aleph_t(\boldsymbol{\varphi}) - \aleph_t(\boldsymbol{\chi})\| \leq \|\boldsymbol{\varphi}_t - \boldsymbol{\chi}_t\| + \|\mathbf{G}_t(\boldsymbol{\varphi})\{Y_t - h_t(\boldsymbol{\varphi})\} - \mathbf{G}_t(\boldsymbol{\chi})\{Y_t - h_t(\boldsymbol{\chi})\}\| \tag{A37}$$

We may finally arrive at the following bound:

$$\|\aleph_t(\boldsymbol{\varphi}) - \aleph_t(\boldsymbol{\chi})\| \leq D_{16} \|\boldsymbol{\varphi}_t - \boldsymbol{\chi}_t\| \tag{A38}$$

where the constant $D_{16} > 0$. Iterating upon this bound (which parallels the derivation of a Taylor-like expansion), we have [39]:

$$\left\| \aleph_{t,k}(\chi) - \aleph_{t,k}(\varphi) \right\|^2 = \int_{t_{i-1}}^{t_i} \left\| \aleph_{t,k}(\chi) - \aleph_{t,k}(\varphi) \right\|_{2,P}^2 dt \leq D_{16}^2 \int_{t_{i-1}}^{t_i} \left\| \aleph_{t,k-1}(\chi) - \aleph_{t,k-1}(\varphi) \right\|_{2,(\nu_{s_1} \times P)}^2 ds_1$$

$$\leq \ldots \leq D_{16}^{2k} \int_{t_{i-1}}^{t_i} \int_{t_{i-1}}^{s_1} \ldots \int_{t_{i-1}}^{s_{k-1}} \left\| \chi - \varphi \right\|^2 ds_k \ldots ds_1$$

$$\leq \frac{D_{16}^{2k} (\Delta t_i)^k}{k!} \left\| \chi - \varphi \right\|^2$$

(A38)

Specifically:

$$\sum_{k=0}^{\infty} \left\| \aleph_{t,k+1}(\varphi_0) - \aleph_{t,k}(\varphi_0) \right\|^2 \leq \left\| \aleph_{t,1}(\varphi_0) - \varphi_0 \right\|^2 \sum_{k=0}^{\infty} \frac{D_{16}^{2k} (\Delta t_i)^k}{k!} < \infty \qquad (A39)$$

From (A39), we see that $\{\aleph_{t,k}(\varphi_0)\}$ is indeed a Cauchy sequence in $L^2(\nu_{t_i} \times P)$.

□

*Proof of Theorem 3:*

Let us consider two solutions $\chi_t$ and $\varphi_t$, whose realizations are continuous paths, arrived at by the proposed iterative algorithm. If $\chi_t = \varphi_t$ a.s. under the product measure $\nu_{t_i} \times P$, then $\chi_t$ and $\varphi_t$ are indistinguishable $P$ a.s. Assume that $\chi_t, \varphi_t \in L^2(\nu_{t_i} \times P)$. As in the last proof, we continue to replace $\|\cdot\|_{2,(\nu_{t_k} \times P)}$ by $\|\cdot\|$. Following (A38), we already have:

$$\left\| \chi - \varphi \right\|^2 = \left\| \aleph_{t,k}(\chi) - \aleph_{t,k}(\varphi) \right\|^2 \leq \frac{D_{16}^{2k} (\Delta t_i)^k}{k!} \left\| \chi - \varphi \right\|^2 \qquad (A40)$$

It is evident from (A40), $\left\| \chi - \varphi \right\| \to 0$ with $k \to \infty$. This implies $\chi_t = \varphi_t$ $(\nu_{t_i} \times P)$ a.s. One muat however show that $\chi_t$, given $\chi_0 = \varphi_0 \in L^2(P)$, belongs to $L^2(\nu_{t_i} \times P)$. Once this is established, the proof is complete. Using Ito's formula:

$$\|\chi_t\|^2 = \|\chi_{i-1}\|^2 + \int_{t_{i-1}}^{t} \left( 2\chi_s^T \chi_s' b(X_s,s) + \|\chi_s'' f(X_s,s)\|^2 \right) ds + \int_{t_{i-1}}^{t} 2\chi_s^T \chi_s' f(X_s,s) dB_s \qquad (A41)$$

Taking expectations on both sides:

$$E_P\left[\|\chi_t\|^2\right] = E_P\left[\|\chi_{i-1}\|^2\right] + E_P\left[\int_{t_{i-1}}^{t} \left( 2\chi_s^T \chi_s' b(X_s,s) + \|\chi_s'' f(X_s,s)\|^2 \right) ds\right] + E_P\left[\int_{t_{i-1}}^{t} \left(2\chi_s^T \chi_s' f(X_s,s)\right) dB_s\right] \qquad (A42)$$

Using linear growth property and noting that the Ito integral $E_P\left[\int_{t_{i-1}}^{t} \left(2\chi_s^T \chi_s' f(X_s,s)\right) dB_s\right] = 0$, we have:

$$E_P\left[\|\chi_t\|^2\right] \leq E_P\left[\|\chi_{i-1}\|^2\right] + D_{17} E_P\left[\int_{t_{i-1}}^{t} \left( 2\|\chi_s\|(1+\|\chi_s\|) + (1+\|\chi_s\|)^2 \right) ds\right] \qquad (A43)$$

where $D_{17} > 0$. (A43) implies:

$$E_P\left[\|\chi_t\|^2\right] \leq E_P\left[\|\chi_{i-1}\|^2\right] + D_{17} E_P\left[\int_{t_{i-1}}^{t} \left((1+\|\chi_s\|) + \|\chi_s\|\right)^2 ds\right] \qquad (A44)$$

(A44) implies:

$$E_P\left[\|\chi_t\|^2\right] \leq E_P\left[\|\chi_{i-1}\|^2\right] + D_{18} E_P\left[\int_{t_{i-1}}^{t} (1+\|\chi_s\|)^2 ds\right] \qquad (A45)$$

where $D_{18} > 0$. Applying Minkowski's inequality and Fubini's theorem:

$$E_P\left[\|\chi_t\|^2\right] \leq E_P\left[\|\chi_{i-1}\|^2\right] + D_{19} \int_{t_{i-1}}^{t} E_P\left[1+\|\chi_s\|^2\right] ds \qquad (A46)$$

where $D_{19} > 0$. Finally applying Gronwall's lemma:

$$E_P\left[\|\chi_t\|^2\right] \leq E_P\left[\|\chi_{i-1}\|^2\right] + E_P\left[1+\|\chi_{i-1}\|^2\right] \exp(D_{19} \Delta t) \qquad (A47)$$

From (A47), the claim follows.

□